\pdfoutput=1

\PassOptionsToPackage{labelfont=bf,labelsep=space}{caption}
\newcommand{\myPaperFontSize}{9pt}
\documentclass[a4paper, \myPaperFontSize, twocolumn, fleqn]{article}

\usepackage{amssymb}
\usepackage[american]{babel}

\usepackage[comma,authoryear,compress]{natbib}		

\usepackage{graphicx}
\graphicspath{{figures/}{}}

\usepackage{amsmath}					
\usepackage[tight,nice]{units}			
\usepackage[babel=true]{csquotes}				
\usepackage{epstopdf}			
\usepackage{subfig}						
\usepackage[hyperfootnotes=false]{hyperref}		
\usepackage[all]{hypcap}
\usepackage{xcolor}
\definecolor{dark-red}{rgb}{0.45,0.15,0.15}
\definecolor{dark-blue}{rgb}{0.15,0.15,0.4}
\definecolor{medium-blue}{rgb}{0,0,0.5}
\hypersetup{
    colorlinks, 				
    linkcolor={dark-red},		
    citecolor={dark-blue},		
    urlcolor={medium-blue},		
}
\usepackage{tabularx}							
\usepackage{booktabs}							
\usepackage[title]{appendix}					
\usepackage{setspace}							
\usepackage{empheq}								
\usepackage[affil-it]{authblk}					

\usepackage[labelfont=bf,labelsep=period, font=small]{caption}

\usepackage{flushend}							

\usepackage{sectsty}							
\sectionfont{\normalsize\selectfont}
\subsectionfont{\normalsize\mdseries\itshape\selectfont}

\usepackage{calc}
\newlength{\mylength}

\renewcommand*{\div}{\mathop{\mathrm{div}}\nolimits}
\DeclareMathOperator{\e}{e}
\newcommand*{\fm}[1]{(\ref{#1})}

\newcommand*{\partialFrac}[2]{{\frac{\partial{#1}}{\partial #2}}}

\newcommand{\protectedSubref}[1]{\protect\subref{#1}}

\newcommand{\cBnd}{{c_\mathrm{bnd}}}
\newcommand{\cMax}{{c_\mathrm{max}}}

\newcommand{\cPl}{{c_\mathrm{pl}}}

\newcommand{\cTot}{{c_\mathrm{tot}}}
\newcommand{\FZeroPerf}{{\widetilde F_\mathrm{0}}}

\newcommand{\FIn}{{F_\mathrm{in}}}
\newcommand{\FInfty}{{F_\infty}}
\newcommand{\FInPerf}{{\widetilde F_\mathrm{in}}}
\newcommand{\FMax}{{F_\mathrm{max}}}
\newcommand{\FPerf}{{\widetilde F}}

\newcommand{\FApprox}{{F^\mathrm{ap}}}
\newcommand{\FApproxMax}{{F^\mathrm{ap}_\mathrm{max}}}

\newcommand{\jCTot}{{\vec j_\cTot}}

\newcommand{\kHenry}{{k_\mathrm{hn}}}
\newcommand{\kHill}{{k_\mathrm{hl}}}
\newcommand{\lFet}{{P}}

\newcommand{\nMF}{{n}}

\newcommand{\nZ}{{\vec n_\mathrm{z}}}

\newcommand{\pONew}{{p_\mathrm{O_2}}}

\newcommand{\PAbs}{{\lFet}}
\newcommand{\PTot}{{P_\mathrm{tot}}}

\newcommand{\RNum}{{R_\mathrm{num}}}
\newcommand{\RNumSquared}{{R_\mathrm{num}^2}}
\newcommand{\rEff}{{r_\mathrm{e}}}

\newcommand{\sFet}{{S_\mathrm{vil}}}

\newcommand{\SM}{{S_\mathrm{IVS}}}

\newcommand{\sTot}{{S_\mathrm{{tot}}}}

\newcommand{\bSixty}{\beta_{60}}
\newcommand{\etaE}{{\eta}}

\newcommand{\gammaPerf}{{\widetilde\gamma}}
\newcommand{\phiOpt}{{\phi_\mathrm{opt}}}

\newcommand{\rhoBl}{{\rho_\mathrm{bl}}}
\newcommand{\tauD}{{\tau_\mathrm{D}}}
\newcommand{\tauEff}{{\tau_\mathrm{e}}}
\newcommand{\tauP}{{\tau_\mathrm{p}}}
\newcommand{\tauPas}{{\tau_\mathrm{tr}}}
\newcommand{\tauTransit}{{\tauPas}}

\addto\captionsenglish{}

\title{\huge \bfseries Analytical theory of oxygen transport in the human placenta}

\author[1]{A.S.~Serov
\thanks{Corresponding author. E-mail: \texttt{alexander.serov@polytechnique.edu}. Corresponding address: Laboratoire de Physique de la Mati\`ere Condens\'ee, Ecole Polytechnique, CNRS, 91128 Palaiseau Cedex, France. Tel.: +33 1 69 33 47 07, Fax: +33 1 69 33 47 99.}
}
\author[2]{C.M.~Salafia}
\author[1]{M.~Filoche}
\author[1]{D.S.~Grebenkov}

\affil[1]{Laboratoire de Physique de la Mati\`ere Condens\'ee, Ecole Polytechnique, CNRS, 91128 Palaiseau Cedex, France}
\affil[2]{Placental Analytics LLC, 93 Colonial Avenue, Larchmont, New York 10538, USA}

\date{\today}

\begin{document}

\maketitle

\begin{abstract}
We propose an analytical approach to solving the diffusion-convection equations governing oxygen transport in the human placenta. 
We show that only two geometrical characteristics of a placental cross-section, villi density and the effective villi radius, are needed to predict fetal oxygen uptake. 
We also identify two combinations of physiological parameters that determine oxygen uptake in a given placenta:~(i) 
the maximal oxygen inflow of a placentone if there were no tissue blocking the flow,
and (ii)~the ratio of transit time of maternal blood through the intervillous space to oxygen extraction time. 
We derive analytical formulas for fast and simple calculation of oxygen uptake and provide two diagrams of efficiency of oxygen transport in an arbitrary placental cross-section. 
We finally show that artificial perfusion experiments with no-hemoglobin blood tend to give a two-orders-of-magnitude underestimation of the \emph{in vivo} oxygen uptake and that
the optimal geometry for such setup alters significantly. The theory allows one to adjust the results of artificial placenta perfusion experiments to account for oxygen-hemoglobin dissociation.
Combined with image analysis techniques, the presented model can give an easy-to-use tool for prediction of the human placenta efficiency.

\vspace{10pt}


\vspace{10pt}

This article was published in the \emph{Journal of Theoretical Biology} \citep{Serov2015Theory} and can be accessed by its doi: \href{http://dx.doi.org/10.1016/j.jtbi.2014.12.016}{10.1016/j.jtbi.2014.12.016}.

\end{abstract}




\section{Introduction}



\begin{figure*}[tb]
\newcommand{\picWidth}{0.37}
\vspace{-4ex} \centering 
\subfloat[]{
\includegraphics[width=\picWidth\paperwidth, natheight=871, natwidth=1063]
{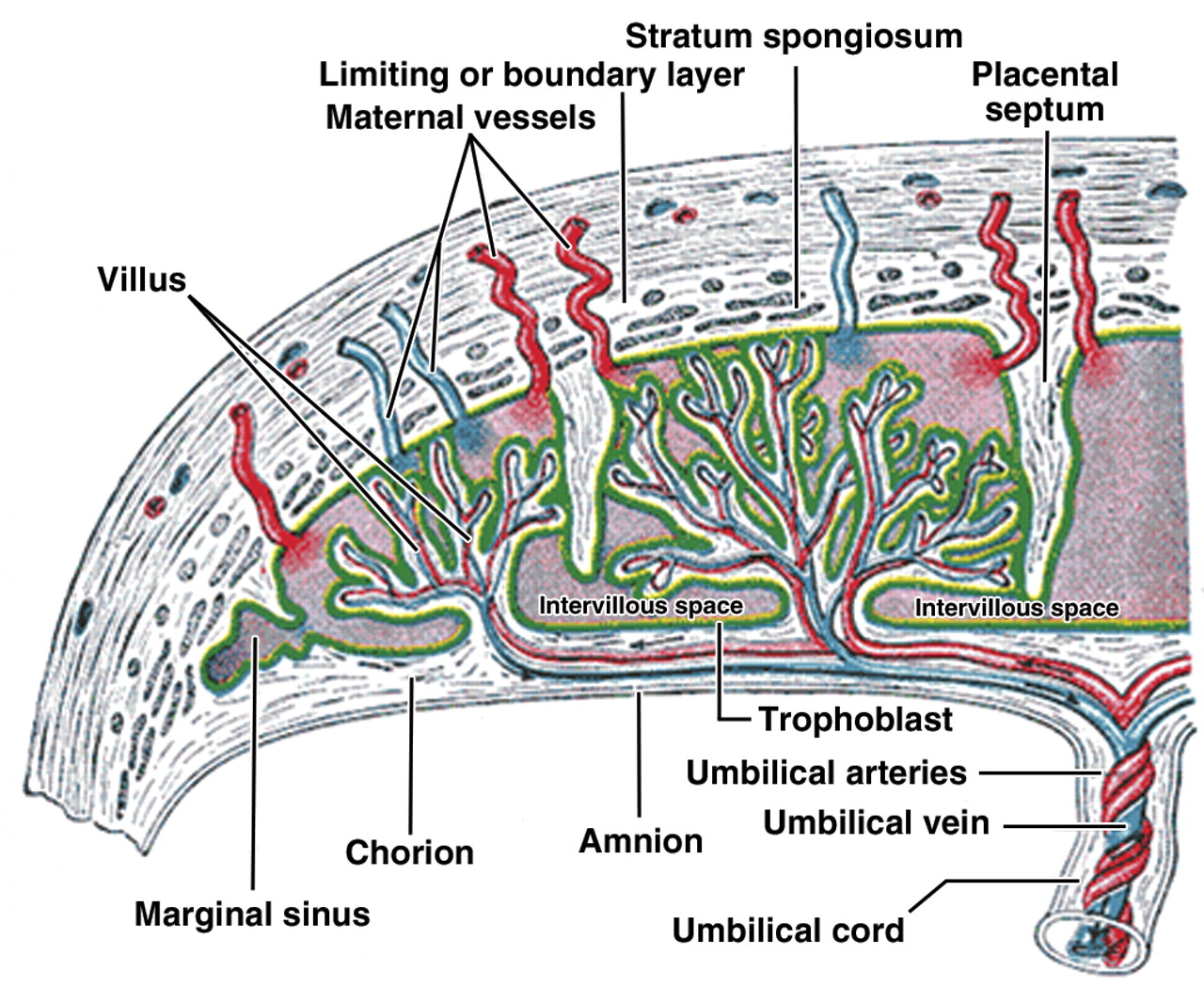}
\label{fig:SchemeOfThePlacenta} }
\hspace{1ex}
\subfloat[]{
\includegraphics[width=\picWidth\paperwidth, natheight=800, natwidth=1096]
{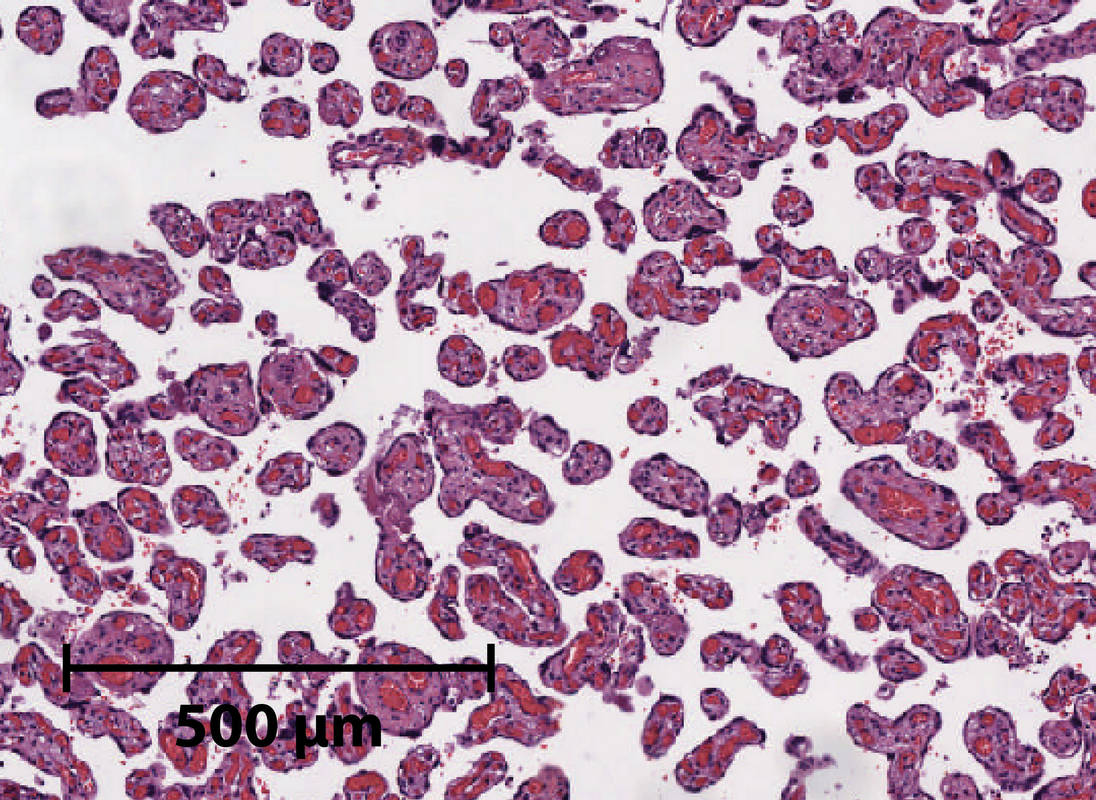}
\label{fig:HealthyPlacentaSlide} }

\caption{
\protectedSubref{fig:SchemeOfThePlacenta}: 
Schematic representation of the human placenta~\citep[reproduced from][]{Gray1918}. Basal plate~(maternal side) is at the top, chorionic plate~(fetal side) is at the bottom.
\protectedSubref{fig:HealthyPlacentaSlide}:~A small fragment of a typical~\mbox{2D} slide of the human placenta. White space is intervillous space, normally filled with maternal blood, which has been washed away during the preparation of the slides (some residual red blood cells are still present). Red shapes are cross-sections of fetal villi. Redder regions inside correspond to fetal capillaries and the dark, violet dots at the perimeter are syncytiotrophoblast boundary layers. The sections have been taken in the direction from the basal plate to the chorionic plate, and are H\&E stained
}
\label{fig:SchemeAndSlideOfThePlacenta}
\end{figure*}

The human placenta consists of maternal and fetal parts~(Fig.~\ref{fig:SchemeOfThePlacenta}). The maternal part is a blood basin which is supplied by spiral arteries and drained by maternal veins~\citep{benirschke_pathology}. The fetal part is a villous tree, inside which fetal blood goes from umbilical arteries to the umbilical vein through fetal capillaries. Maternal blood percolates through the same arboreous structure on the outside. Maternal blood and fetal blood do not mix, so the gas and nutrient exchange takes place at the surface of the villous tree, sections of which can be observed in a typical histological~\mbox{2D} placental slide~(Fig.~\ref{fig:HealthyPlacentaSlide}). Modeling and understanding the relation between the geometrical structure of the exchange surface of the villous tree and the efficiency of the transport function of the placenta constitutes the central object of our study. 

Placenta models have been proposed previously~\citep[see discussions in][]{Aifantis1978,
Battaglia1986,
chernyavsky_2010,
Gill2011}.
\mbox{1D} models dealt with oxygen transport at the scale of either one single villus or the whole placenta, in both cases imposing a flat exchange surface between maternal and fetal blood~\citep{
Bartels1962,
Shapiro1967,
Kirschbaum1969,
Hill1972,
Hill1973,
Longo1972a, Longo1972b,
Power1972, Power1972a,
Lardner1975,
Wilbur1978,
Groome1991};
some \mbox{2D} models were used to study the co-orientation of maternal and fetal flows~\citep{
Bartels1962,
Metcalfe1964,
Shapiro1967,
Faber1969,
Kirschbaum1969,
Guilbeau1970,
Moll1972,
Schroder1982b,
Battaglia1986};
other models represented the placenta as a porous medium~\citep{
Erian1977,
Schroder1982b,
chernyavsky_2010}.
A lumped element model was also proposed to calculate~1D placental diffusing capacity and to relate morphometric data to the efficiency of gas transport~(see~\citealp{Mayhew1984,Mayhew1986} and references therein).
To our knowledge, the only \mbox{3D} placenta model was introduced by~\citet{chernyavsky_2010} to study how the position of venous outlets and the existence of a central cavity influences oxygen transport in a hemispherical porous-medium placentone model.

However, none of these models uses fine geometrical structure of experimentally obtained placental slides~(Fig.~\ref{fig:HealthyPlacentaSlide}) as direct input.
In a recent paper we introduced a stream-tube placenta model~(STPM; Fig.~\ref{figPlacentaModel3D}), which is built upon histological placental cross-sections~\citep{Serov2015Optimality} in contrast to previous placenta models. In this model, cross-sections of stream tubes of maternal blood flow~(MBF) in the intervillous space of the human placenta were reconstructed from placental cross-sections~(Fig.~\ref{fig:HealthyPlacentaSlide}) and virtually extended along the third dimension. 
Although successive cross-sections of a stream tube obviously vary in the placenta, this variation cannot be reproduced from a single cross-section and was ignored in this model.
Relevant physiological and geometrical parameters of the model were estimated from available experimental data. 
Numerical simulations of oxygen transport for identical circular villi were then performed and showed that the model exhibits an optimal villi density yielding maximal oxygen uptake. Deviations from these optimal characteristics with variations of model parameters were estimated. The obtained optimal villi density~($0.47\pm0.06$) corresponds to that experimentally obtained in healthy human placentas~($0.46\pm0.06$). %

The present manuscript relies on the same~STPM, but provides the first approximate \emph{analytical} theory of oxygen uptake in the human placenta based on histological cross-sections. The present work significantly develops the results of the previous study by:
\begin{itemize}
\item
allowing for a fast calculation of oxygen uptake for arbitrary placental cross-sections, while only circular villi were considered before;

\item
demonstrating explicit dependence of oxygen uptake on model parameters and their interrelation, which could not be obtained numerically; 

\item
introducing two uptake efficiency indicators for which analytical formulas and diagrams are provided;

\item
showing that accounting for oxygen--hemoglobin reaction is important for interpretation of artificial perfusion experiments~(with no-Hb blood) and providing a method of recalculation of the results of such experiments to account for oxygen--hemoglobin reaction.
\end{itemize}
In the following, the construction of the analytical theory is preceded by a full description of the geometrical model, physical assumptions and parameters describing the human placenta. 


\section{The model}
\subsection{Model assumptions}

\begin{figure*}[tb]
\newcommand{\picWidth}{0.37}
\vspace{-4ex} \centering 
\subfloat[]{
\includegraphics[width=\picWidth\paperwidth, natheight=800, natwidth=1096]
{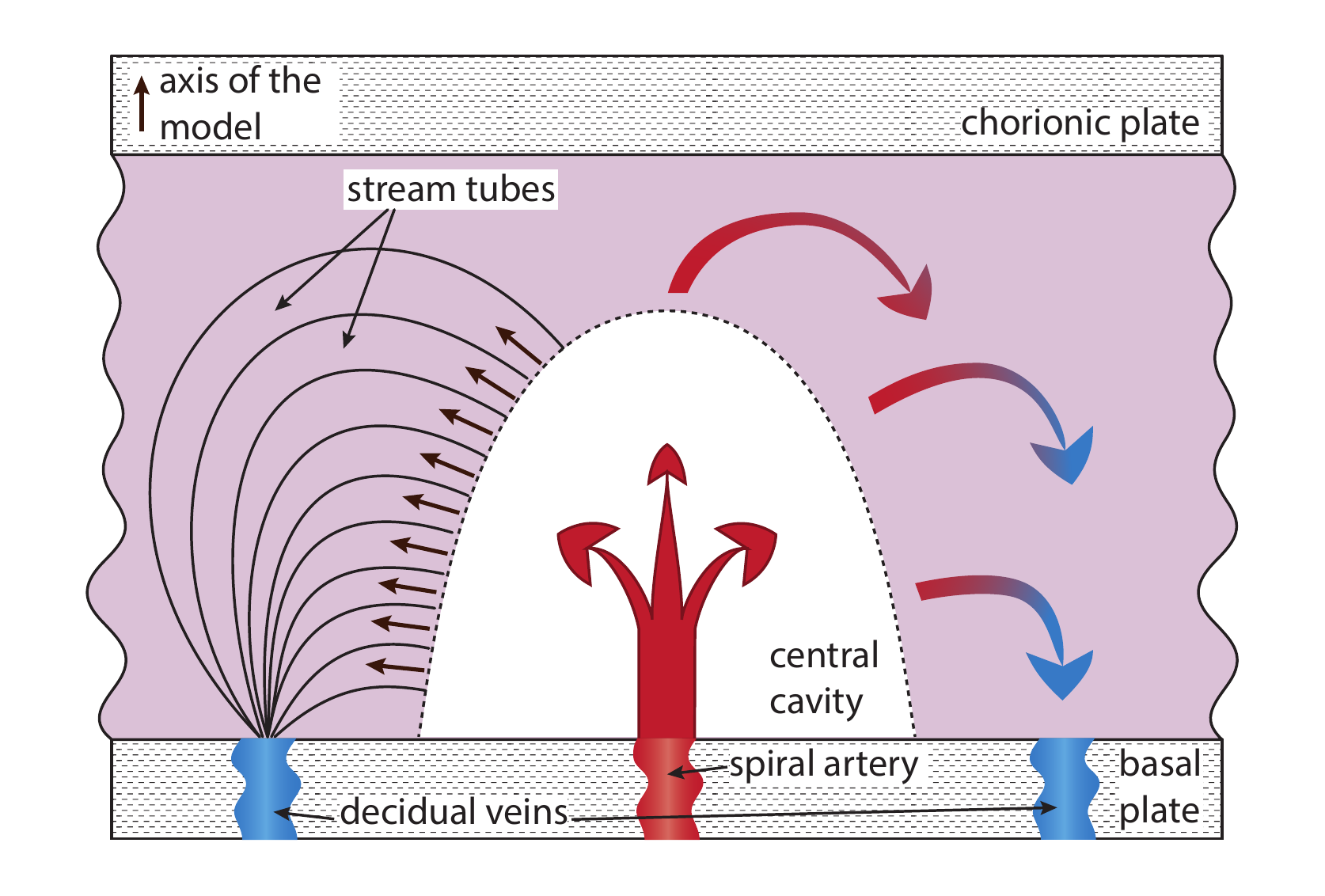}
\label{fig:positioning_of_the_geometry} }
\hspace{1ex}
\subfloat[]{
\includegraphics[width=0.3\paperwidth, natheight=477, natwidth=912]
{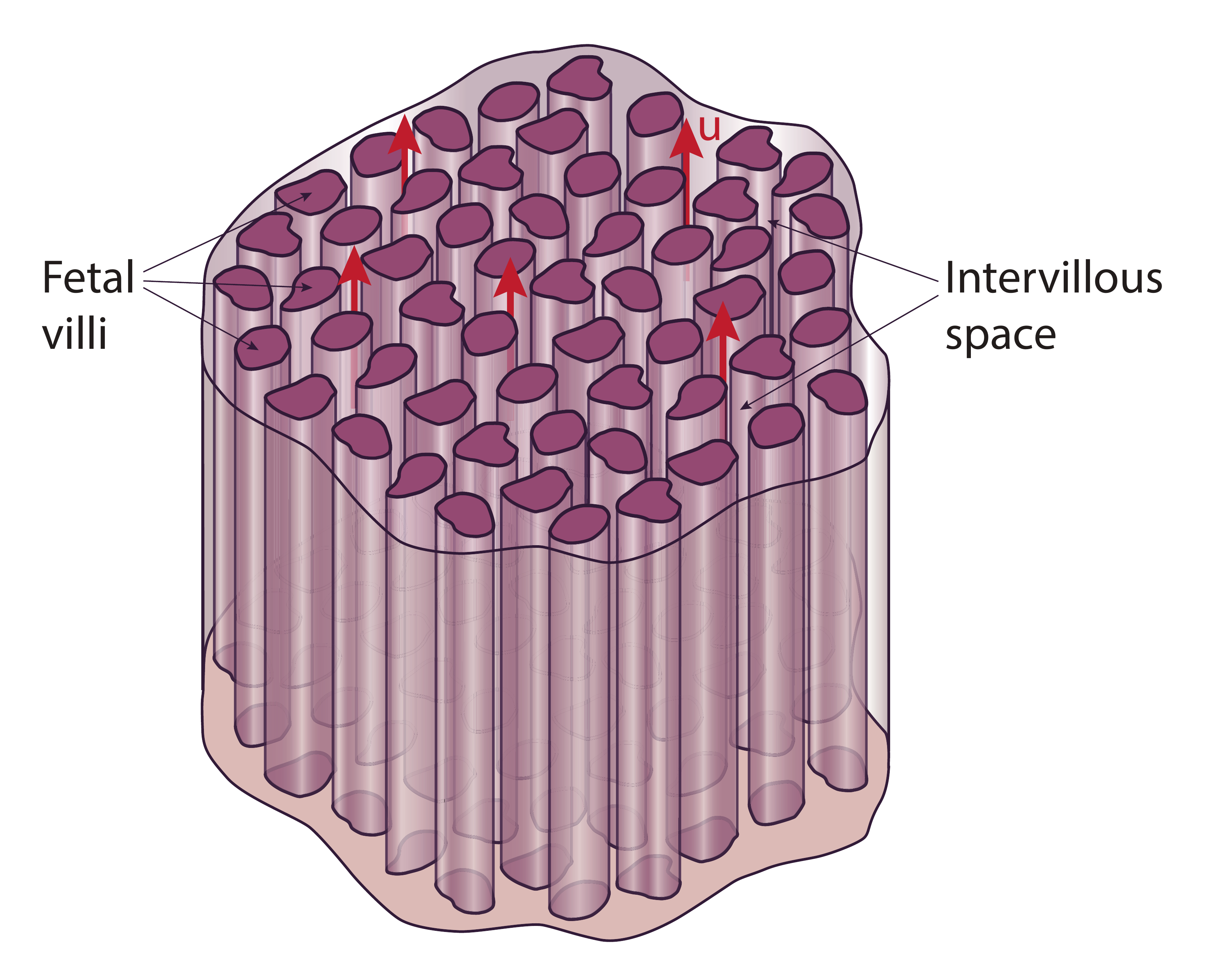}
\label{figPlacentaModel3D} }
\caption{
\protectedSubref{fig:positioning_of_the_geometry}~%
Scheme of placental blood flow and location of stream tubes in the placenta.
The dashed line schematically outlines the central cavity. Curved arrows in the right part show maternal blood losing oxygen while going from the central cavity to decidual veins percolating through the villous tree~(shaded region). Curved lines on the left schematically show stream tubes of blood flow. Our model corresponds to one such stream tube unfolded; small straight arrows show the entrance points of the model.
The exchange is not modeled in the central cavity, but only after it, in the~MBF pathway. The concept that spiral arteries open into the~IVS near the central cavity corresponds to current physiological views~\citep{benirschke_pathology,chernyavsky_2010}.
\protectedSubref{figPlacentaModel3D}
Our geometrical model of one unfolded stream tube of maternal blood in a human placentone. The cross-section of the stream tube as well as that of the fetal villi inside can be either determined from~2D histological slides or chosen arbitrary.
Red arrows show the flow of maternal blood
}
\label{fig:3DModelAndPositioning}
\end{figure*}

Maternal blood arrives into the intervillous space of the human placenta by spiral arteries~(Fig.\ref{fig:positioning_of_the_geometry}). 
It then percolates through the branching structure of a tree of fetal villi and leaves the intervillous space by decidual veins. 
The total pattern of the~MBF can be virtually subdivided into small regions~(stream tubes), each following the flow and extending from the central cavity to a decidual vein. 
Each stream tube comes into contact with numerous fetal villi, at the surface of which mass exchange between maternal and fetal blood takes place.

We model one such stream tube \emph{unfolded} as a cylinder of an arbitrary cross-section containing multiple parallel cylinders of arbitrary cross-sections and sizes, which represent fetal villi~(Fig.~\ref{figPlacentaModel3D}).
The shapes and locations of the villi can be taken from a histological slide. %
Since we aim to base the~STPM on histological slides~(Fig.~\ref{fig:HealthyPlacentaSlide}) which provide only one stream-tube section without any information about the change of this section along the~MBF,
we further postulate that the same shapes and locations of villi are conserved along the stream tube. %
This is obviously an oversimplification of the irregular~3D structure of the placenta, but it is the most straightforward assumption given the lack of complete~3D geometrical data spatially resolving all villi. 

The model relies on several other assumptions:
\begin{enumerate}

\item Fetal blood is considered as a perfect oxygen sink.

\item MBF is considered to be laminar with slip conditions at all boundaries~(no liquid-wall friction), so that the velocity profile in any cross-section is flat.

\item The oxygen--hemoglobin dissociation curve is linearized in the physiological range of partial pressure of oxygen~($\unit[{0\text{--}60}]{mmHg}$) observed in the human placenta.

\item Oxygen uptake occurs at the feto-maternal interface, i.e. at the boundaries of the small cylinders, and is directly proportional to the interface permeability and to oxygen concentration on the maternal side of the interface.

\item In a cross-section perpendicular to the MBF, oxygen is only redistributed by diffusion.

\item Erythrocytes are uniformly distributed in the maternal blood.

\item Oxygen bound to hemoglobin does not diffuse; only oxygen dissolved in the blood plasma does.

\item Oxygen uptake is stationary.

\end{enumerate}
The validity of these assumptions is thoroughly discussed in~\citet{Serov2015Optimality}.
The model also includes geometrical and biological parameters, which are listed in Table~\ref{tblCalculationsParameters}. 
It will further be shown that fetal oxygen uptake is determined by two parameter combinations which naturally appear in the development of the theory.


\begin{table*}[tb]
	\caption{\rmfamily
	Parameters of the human placenta used in the~STPM~(see~\citeauthor{Serov2015Optimality},~\citeyear{Serov2015Optimality} for the method of calculation of these values)}
	\smallskip
	\label{tblCalculationsParameters}
	    \centering	
		\begin{tabularx}{\textwidth}{@{}p{4.7in}lr@{\ }l@{}}
			\toprule
			Parameter & Symbol & Mean & $\pm$ SD
			\\
			\midrule
			Maximal Hb-bound oxygen concentration at~\unit[100]{\%} Hb saturation,~$\unit{mol/m^{3}}$	&	$\cMax$	&	7.30 & $\pm$ 0.11
			\\
			Oxygen-hemoglobin dissociation constant	&	$B$	&	94 & $\pm$ 2
			\\
			Concentration of oxygen dissolved in blood at the entrance to the~IVS,~$\unit[10^{-2}]{mol/m^{3}}$	&	$c_0$	&	6.7 & $\pm$ 0.2
			\\			
			Oxygen diffusivity in blood,~$\unit[10^{-9}]{m^2/s}$	&	$D$	&	1.7 & $\pm$ 0.5
			\\			
			Effective villi radius,~$\unit[10^{-6}]{m}$	&	$\rEff$	&	41 & $\pm$ 3
			\\
			Permeability of the effective materno-fetal interface,~$\unit[10^{-4}]{m/s}$	&	$w$	&	2.8 & $\pm$ 1.1
			\\			
			Placentone radius,~$\unit[10^{-2}]{m}$	&	$R$	&	1.6 & $\pm$ 0.4
			\\
			Velocity of the maternal blood flow,~$\unit[10^{-4}]{m/s}$	&	$u$	&	6 &
			\\
			Stream tube length,~$\unit[10^{-2}]{m}$	&	$L$	&	1.6 &
			\\
			\bottomrule
		\end{tabularx}
\end{table*}



\section{Mathematical formulation}
\label{sect:mathematical_formulation}

\subsection{Time scales of the system}


We identify three different physical transport processes in the placenta, each of which operates on a characteristic time scale: hydrodynamic blood flow through the~IVS characterized by an average velocity~$u$ and transit time~$\tauPas$; diffusion of oxygen with characteristic time~$\tauD$; and equilibration between oxygen bound to hemoglobin and oxygen dissolved in the blood plasma with characteristic time~$\tauP$. This last thermodynamic equilibrium is described as equal partial pressure of oxygen in both states.
The three times can be estimated as follows:
\begin{itemize}
\item $\tauPas$, the transit time of blood through the~IVS, is of the order of~\unit[27]{s} from the results of angiographic studies at term~\citep{Burchell1967,Serov2015Optimality}.

\item $\tauD$, reflecting oxygen diffusion over a length~$\delta$ in the~IVS is~%
$\tauD\sim\delta^2/D$, where~$D$ is oxygen diffusivity in the blood plasma~(Table~\ref{tblCalculationsParameters}).
Either from calculations~\citep{Mayhew2000}, or directly from %
normal placental sections~(Fig.~\ref{fig:HealthyPlacentaSlide}), the mean width of an~IVS pore can be estimated as~$\delta\sim\unit[80]{\mu m}$, yielding~$\tauD\sim\unit[4]{s}$.

\item $\tauP$, an equilibration time scale, which includes characteristic diffusion time for oxygen to reach Hb~molecules inside a red blood cell~\citep[{$\sim\unit[10]{ms}$}, see][]{Foucquier2013} and typical time of oxygen--hemoglobin dissociation~($\sim\unit[20]{ms}$ for the slowest process, see~\citealp{Yamaguchi1985}). Together these times sum up to~$\tauP\sim\unit[30]{ms}$.
\end{itemize}
These three characteristic times are related as follows:~$\tauP\ll\tauD\lesssim\tauPas$. This relation suggests that oxygen-hemoglobin dissociation can be considered instantaneous as compared to diffusion and convection; the latter two, by contrast, should be treated simultaneously.

\subsection{Equilibrium between bound and dissolved oxygen}

The very fast oxygen--hemoglobin reaction can be accounted for by assuming that the concentration of oxygen dissolved in the blood plasma~($\cPl$) and that of oxygen bound to hemoglobin~($\cBnd$) instantaneously mirror each other's changes. Mathematically both concentrations can be related by equating oxygen partial pressures in these two forms.

\paragraph{Oxygen dissolved in plasma.}
Because of the low solubility of oxygen in blood, the partial pressure of the dissolved oxygen~($\pONew$) can be related to its concentration~($\cPl$) using Henry's law:
\begin{equation}
\label{fmHenrysLaw}
\pONew=\frac\kHenry \rhoBl \cdot\cPl,
\end{equation}
where~$\rhoBl\approx\unit[1000]{kg/m^3}$ is the density of blood and the coefficient~$\kHenry$ can be estimated from the fact that a concentration~$\cPl\approx\unit[0.13]{mol/m^3}$ of the dissolved oxygen corresponds to oxygen content of~$\unit[3]{ml\ O_2/l\ blood}$ or partial pressure of~$\unit[13]{kPa}$ at normal conditions~\citep{Law1999}, yielding~$\kHenry\sim\unit[7.5\cdot10^{5}]{mmHg\cdot kg/mol}$
for the oxygen dissolved in blood.

\begin{figure}[tbH!]
\center{\includegraphics[width=3in, natwidth=1905, natheight=1254]
{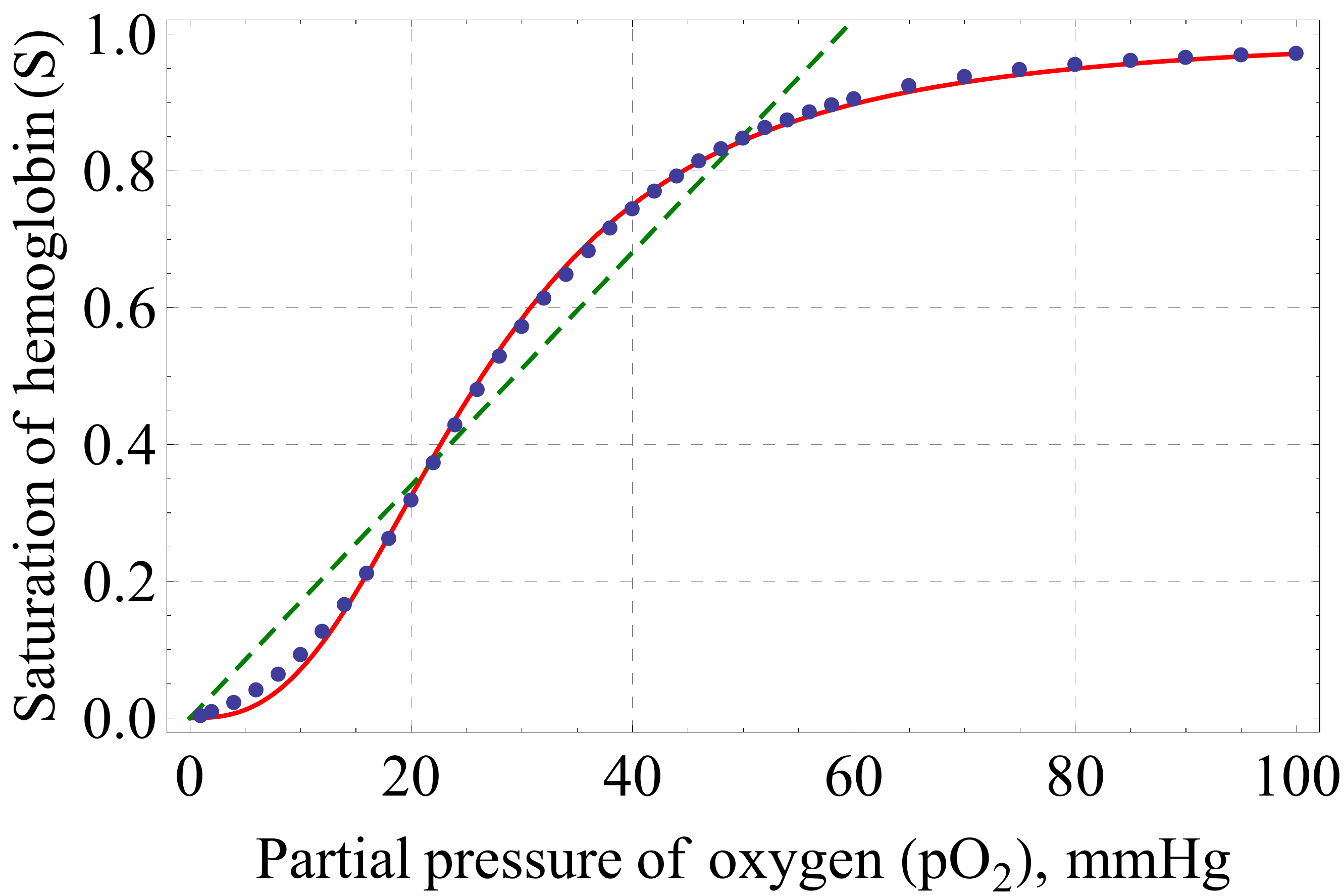}}
\caption{
Oxygen--hemoglobin dissociation curve. In the figure:~(i) dots are experimental data at normal conditions as obtained by~\citet{Severinghaus1979};
(ii) solid curve shows a fit of these data with Hill equation, the coefficients being~$\alpha=2.65$, $\kHill=\unit[0.04]{mmHg^{-1}}$;
(iii) straight dashed line is a linear approximation of the curve in the~$\unit[0\text{--}60]{mmHg}$ region as discussed in Section~\ref{sectLinearizationOfTheHillsLaw}, slope of the line being~$\bSixty\approx\unit[0.017]{mmHg^{-1}}$
}
\label{figHillsLaw}
\end{figure}

\paragraph{Hemoglobin-bound oxygen.}
The partial pressure of the hemoglobin-bound oxygen depends on its concentration through Hill equation:
\begin{align}
\label{fmSaturationDefinitionHillEquation}
&\cBnd=\cMax\ S(\pONew), &&S(\pONew)\equiv \frac{(\kHill\pONew)^\alpha}{1+(\kHill\pONew)^\alpha},
\end{align}
where~$\cMax$ is the oxygen content of maternal blood at full saturation;~$\kHill\approx\unit[0.04]{mmHg^{-1}}$ and \mbox{$\alpha\approx2.65$} are coefficients of Hill equation, obtained by fitting the experimental curve of~\citet[see Fig.~\ref{figHillsLaw}]{Severinghaus1979}.

Equilibrium relation between~$\cPl$ and~$\cBnd$ can then be obtained by substituting Eq.~\fm{fmHenrysLaw} into Eq.~\fm{fmSaturationDefinitionHillEquation}:
\begin{equation}
\label{fmMirrorRelation}
\cBnd=\cMax\ S\left(\frac\kHenry\rhoBl \cPl\right).
\end{equation}

\subsection{Diffusive-convective transport of oxygen}

Diffusive-convective transport of oxygen is governed by the mass conservation law for the total concentration of oxygen in a volume of blood:
\begin{equation}
\label{fmMassConservationLaw}
\partialFrac{(\cPl+\cBnd)}{t}+\div \jCTot=0,
\end{equation}
where~$\jCTot$ is the total flux of oxygen, transported both by diffusion and convection for the dissolved form and only by convection~(RBCs being too large objects) for the bound form:
\begin{equation}
\label{fmJCTotExpression}
\jCTot=-D\vec\nabla\cPl+\vec u(\cPl+\cBnd),
\end{equation}
where $\vec u$ denotes the velocity of the MBF and
\begin{equation*}
\overrightarrow\nabla\equiv \left\{\frac \partial {\partial x}; \frac \partial {\partial y}; \frac \partial {\partial z}\right\}.
\end{equation*}
Omitting the time derivative in the stationary regime,  substituting Eq.~\fm{fmJCTotExpression} into Eq.~\fm{fmMassConservationLaw} and choosing~$z$ as the direction of the~MBF, we obtain
\[
\Delta\cPl=\frac uD \partialFrac{(\cPl+\cBnd)}{z},
\]
where 
\[
\Delta\equiv \partialFrac{^2}{x^2}+\partialFrac{^2}{y^2}+\partialFrac{^2}{z^2}
\]
is the Laplace operator.
Using the relation~\fm{fmMirrorRelation} between the dissolved and bound oxygen concentrations we then derive an equation for the unknown~$\cPl$ only:
\begin{equation}
\label{fmMainEquationNonLinear}
\begin{aligned}
\Delta\cPl&=\frac u D\partialFrac{ }{z}\left(\cPl+\cMax\ S\left(\frac \kHenry \rhoBl \cPl\right)\right)
\\
&=\frac u D \left(1+\frac{\cMax \kHenry}{\rhoBl}\ S'\left(\frac \kHenry \rhoBl \cPl\right)\right)\partialFrac{\cPl}{z}.
\end{aligned}
\end{equation}
This equation is non-linear as $\cPl$ appears also in the argument of the derivative of Hill saturation function~$S'$. In a first approximation,~$S$ can be linearized by assuming~$S'$ to be constant in the range of partial pressures of oxygen encountered in the human placenta.

\subsection{Linearization of Hill equation}
\label{sectLinearizationOfTheHillsLaw}

The idea of linearization is simple: to replace the sigmoid saturation function~\fm{fmSaturationDefinitionHillEquation} with a linear function of~$\pONew$. 
Although it is natural to make the line pass through the origin,
the slope of the line may be chosen differently depending on the range of partial pressures in which we approximate the curve~(Fig.~\ref{figHillsLaw}). Data found in the literature
indicate that maternal blood \emph{in the~IVS} of the human placenta has~$\pONew$ of about~$\unit[60]{mmHg}$~\citep{Rodesch1992, Jauniaux2000, Challier2003}. We further suppose that this partial pressure is the maximal value in the~IVS and hence delimits the range of the needed linear approximation. Fitting the experimental curve of~\citet{Severinghaus1979} in the region~$\unit[0\text{--}60]{mmHg}$ with a straight line passing through zero we obtain a linear approximation
\begin{align*}
&S(\pONew)\approx\bSixty\, \pONew, 	&&S'(\pONew)=\bSixty\approx\unit[0.017]{mmHg^{-1}},
\end{align*}
displayed in Fig.~\ref{figHillsLaw}.

This approximation leads to the following relation between~$\cTot$,~$\cPl$ and~$\cBnd$:~$\cTot\equiv\cPl+\cBnd=\cPl B$,
or
\begin{align*}
&\cPl=\frac \cBnd{B-1} =\frac {\cMax}{B-1}S(\pONew),
\\
&\text{where\ }
B\equiv1+\frac{\cMax\bSixty\kHenry}{\rhoBl}.
\end{align*}
We emphasize here that ignoring oxygen-hemoglobin interaction would be equivalent to setting $B=1$, which would lead to a hundred-fold underestimation of this constant~(Table~\ref{tblCalculationsParameters}).

From Eq.~\fm{fmJCTotExpression}, a linearized version of the corresponding total oxygen flux is then
\begin{equation}
\label{fmJCTotExpressionWithB}
\jCTot=-D\vec\nabla \cPl +\vec u\cPl B.
\end{equation}

Finally the partial differential equation~\fm{fmMainEquationNonLinear} becomes
\begin{equation}
\label{fmGeneralEquationLinear}
\Delta\cPl=\frac {uB}D\frac{\partial \cPl}{\partial z}.
\end{equation}

\subsection{Boundary conditions}

Boundary conditions should be imposed on Eq.~\fm{fmGeneralEquationLinear}:
\begin{itemize}
\item the boundary of the large cylinder represents the outer boundary of a stream tube. Assuming there is no exchange of oxygen between different stream tubes%
, we consider zero flux on its wall:
\[
\partialFrac{\cPl}{\nMF}=0,
\]
where~$\partial/\partial\nMF$ is the normal derivative directed outside the~IVS;

\item uptake at the effective feto-maternal interface is proportional to the concentration of oxygen dissolved in the maternal blood plasma:
\begin{equation*}
D\partialFrac{\cPl}{n}+w\cPl=0,
\end{equation*}
where~$w$ is the permeability of the interface which accounts for the resistance of IVS--villus and villus--capillary membranes as well as for diffusion in the connective tissue separating the two membranes;

\item the total concentration of oxygen in blood is uniform and constant 
at the entrance of the stream tube~($z=0$):
\begin{align*}
&\cPl(x,y,z=0)=c_0, &\forall (x,y)\in \SM,
\end{align*}
where~$c_0$ is oxygen concentration in the incoming blood plasma and~$\SM$ is the part of the stream-tube cross-section occupied by the~IVS.

\end{itemize}


\subsection{Conversion to a~\mbox{2D} eigenvalue equation}

To solve Eq.~\fm{fmGeneralEquationLinear}, we separate the coordinate~$z$ along the stream-tube axis from the coordinates~$x$ and~$y$ in the transverse cross-section.
The general solution of Eq.~\fm{fmGeneralEquationLinear} then takes the following form:
\begin{equation}
\label{fmSolutionForm}
\cPl(x,y,z)=c_0\sum_{j=1}^\infty a_j v_j(x,y)\e^{-\mu_jz},
\end{equation}
where~%
$\{\mu_j\}$ are decay rates in the~$z$-direction and~$\{a_j\}$ are weights of~$\cPl$ in the orthonormal eigenbasis~$\{v_j(x,y)\}$ of the Laplace operator~$\Delta_{xy}$ in the transverse cross-section. $\{v_j(x,y)\}$ satisfy the following equations:

\setlength{\mylength}{-1ex}
{
\vspace{\mylength}
\small
\begin{empheq}[left=\empheqlbrace]{align}
\label{fmEigenValueEquation01}
&(\Delta_{xy}+\Lambda_j)\ v_j=0,&&
\\
\label{fmEigenValueEquation02}
&\partialFrac{v_j}{\nMF}=0&&\text{on stream-tube boundary},
\\
\label{fmEigenValueEquation03}
&\left(\partialFrac{}{\nMF}+\frac wD\right)v_j=0&&\text{on villi boundaries},
\\
\label{fmEigenValueEquation04}
&\sum\limits_j a_j v_j=1&&\text{in the $z=0$ plane},
\end{empheq}
}

\setlength{\mylength}{-5ex}
\vspace{\mylength}
\begin{flalign}
&\text{where}
&&\Delta_{xy}\equiv\frac{\partial^2}{\partial x^2}+\frac{\partial^2}{\partial y^2},
&&\Lambda_j\equiv \mu_j^2+\mu_j\frac {uB}D.&&
\label{fm:LambdaDefinition}
\end{flalign}%
Eigenvalues~$\{\mu_j\}$, eigenfunctions~$\{v_j(x,y)\}$ and weights~$\{a_j\}$ are determined by Eqs~\fm{fmEigenValueEquation01}--\fm{fm:LambdaDefinition} for a given cross-section. In particular, from Eq.~\fm{fmEigenValueEquation04} it follows that~%
$a_j\equiv\int_\SM v_jdS$,
when eigenfunctions are $L_2$-normalized~($\int_\SM v_j^2=1$).

\subsection{General expression for oxygen uptake}
According to the mass conservation law, oxygen uptake up to length~$L$ is equal to the difference between oxygen flow coming into the system at~$z=0$ and oxygen flow leaving the system at~$z=L$. Using Eqs~\fm{fmJCTotExpressionWithB} and~\fm{fmSolutionForm} %
one can derive an explicit dependence of oxygen uptake on the stream-tube length:
\begin{align}
\notag
F(L)&=\int\limits_\SM \left.\left(\jCTot\cdot\nZ\right)\right|_{z=0}dS-
\int\limits_\SM \left.\left(\jCTot\cdot\nZ\right)\right|_{z=L}dS
\\
\label{fmFinalGeneralExpressionForF}
&=c_0\sum\limits_{j=1}^\infty a_j^2\left(D\mu_j+uB\right)\left(1-\e^{-\mu_jL}\right),
\end{align}
where the definition of~$\{a_j\}$ has been used.
This is an exact expression for oxygen uptake, into which the geometrical structure of the placental cross-section enters through the spectral characteristics~$\{\mu_j\}$ and~$\{a_j\}$. Our goal now is to simplify this expression and to identify the most relevant geometrical and physiological parameters that determine oxygen uptake.

\section{Approximate analytical solution}

\subsection{Form of the approximation}

A quick analysis of Eq.~\fm{fmFinalGeneralExpressionForF} shows that~$F(L)$ is a smooth monotonous curve which is linear at small lengths and exponentially saturates at large lengths. In a first approximation, Eq.~\fm{fmFinalGeneralExpressionForF} can then be replaced by an expression which has the same behavior at these limits:
\begin{equation}
\label{fmApproximateFExpression}
\FApprox(L)=A(1-\e^{-\alpha L}),
\end{equation}
where~$A$ and~$\alpha$ are two parameters:~$A$ is oxygen uptake at~$L\to\infty$~(equal to the total incoming oxygen flow), and~$\alpha$ is the mean decay rate of oxygen concentration with stream-tube length.
We will now relate~$A$ and~$\alpha$ to the parameters of the model.

\subsection{Uptake at the infinite length}

Large lengths are characterized by saturation when all incoming oxygen is transferred to the fetal blood. 

An exact expression for the saturation limit can be obtained from Eq.~\fm{fmFinalGeneralExpressionForF} as~$L\to\infty$:
\begin{equation}
\label{fmFInftyTwoComponents}
\FInfty=c_0uB\SM+c_0D \sum\limits_{j=1}^\infty a_j^2\mu_j,
\end{equation}
where the identity~$\sum_{j=1}^\infty a_j^2=\SM$, resulting from Eq.~\fm{fmEigenValueEquation04}, was used. 


Note that the flow in Eq.~\fm{fmFInftyTwoComponents} includes two contributions: convective flow (the first term) and diffusive flow along the~$z$-axis~(the second term). It turns out that the second term is much smaller than the first one, so that Eq.~\fm{fmFInftyTwoComponents} can be simplified by omitting the diffusive term:
\begin{equation}
\label{fm:parameter_A}
\FInfty\approx A=c_0uB\SM.
\end{equation}
This approximation is justified by the following arguments:
\begin{enumerate}
\item 
Relative roles of convection and diffusion in a hydrodynamic problem are described by the P\'eclet number, which is the ratio of characteristic times of diffusive and convective transport to the same distance~$\delta$:~$\mathrm{Pe}=u\delta/D$.
Large P\'eclet numbers~($\mathrm{Pe}\gg1$) indicate predominant convective transport, whilst small values signify prevalence of the diffusive transport.

For the human placenta, the ratio~$u/D$ is of the order of~$\unit[10^5]{m^{-1}}$~(see Table~\ref{tblCalculationsParameters}), which yields that~$\mathrm{Pe}\gg1$ for lengths~$\delta\gg\unit[10]{\mu m}$. As the characteristic length of the stream tube is~$L_0\sim\unit[1.6]{cm}\gg\unit[10]{\mu m}$, we conclude that diffusion along the stream tube can be omitted as compared to convection. At the same time, diffusion in the cross-section cannot be ignored as it is the only in-plane mechanism of oxygen transport.

\item Since~\unit[99]{\%} of oxygen is bound to hemoglobin in red blood cells~(RBC), and~RBCs are too large to diffuse, the error from ignoring the diffusive transport term in~$\FInfty$ does not exceed~\unit[1]{\%} in terms of oxygen content.

\end{enumerate}


Mathematically, the simplification we have used can be written as~
\begin{equation}
\mu_j\ll uB/D,
\label{fm:LinearMuApproximation}
\end{equation}
so that Eq.~\fm{fm:LambdaDefinition} becomes~$\mu_j\approx\Lambda_j D/(uB)$.
It should be noted that statement~\fm{fm:LinearMuApproximation} does not contradict with the fact that~$\{\mu_j\}$ grow to infinity with eigenvalue number~$j$. In fact, each~$\mu_j$ contributes to the final expression with a weight~$a_j$, which diminishes with~$j$. Eq.~\fm{fm:LinearMuApproximation} should be then understood as only valid for all eigenvalues that have significant contributions~$a_j$.


Oxygen uptake~\fm{fmFinalGeneralExpressionForF} can then be approximated as
\begin{equation}
F(L)
\approx c_0uB\SM\left(1-\sum\limits_{j=1}^\infty \frac{a_j^2}\SM\exp\left(-\frac D{uB}\Lambda_jL\right)\right).
\label{fm:FWithLambda}
\end{equation}

\subsection{Average concentration decay rate}

Using Eq.~\fm{fm:parameter_A} and comparing Eq.~\fm{fm:FWithLambda} with the approximate form of oxygen uptake~\fm{fmApproximateFExpression}, one obtains the following definition of the average concentration decay rate~$\alpha$:
\begin{equation}
\alpha(L)\equiv-\frac1L \ln\left(\sum\limits_{j=1}^\infty \frac{a_j^2}\SM\exp\left(-\frac D{uB}\Lambda_jL\right)\right).
\label{fm:alphaFirstExpression}
\end{equation}
In this formula,~$\alpha$ depends explicitly on~$L$ and implicitly on the cross-sectional geometry. Our goal now is to extract the main part of this implicit dependence. For this purpose, we integrate Eq.~\fm{fmEigenValueEquation01} over the~IVS in the cross-section~($\SM$). We further apply the divergence theorem~\citep{Arfken2005} to transform the integral over the~IVS to an integral over its boundary~($\PTot$, which includes the perimeter~$P$ of the absorbing boundary of the villi and that of the outer boundary of the stream tube in a cross-section):
\begin{align}
\Lambda_j&=-\frac{\int_\PTot \partial v_j/\partial n\ dP}{\int_{\SM} v_j\ dS}
\nonumber
\\
&=\frac wD\frac{\int_\PAbs v_j\ dP}{\int_{\SM} v_j\ dS}
=\left(\frac {w\PAbs}{D\SM}\right)q_j^2,
\label{fm:LambdaProperty}
\end{align}
where boundary conditions~\fm{fmEigenValueEquation02} and~\fm{fmEigenValueEquation03} were used to remove the contribution from the non-absorbing boundary and
\begin{equation*}
q_j^2\equiv \frac{\frac 1\PAbs\int_\PAbs v_j\ dP}{\frac 1\SM\int_{\SM} v_j\ dS}%
\end{equation*}
is the dimensionless ratio of the mean value of the eigenfunction~$v_j$ over the villi boundary to its mean value in the~IVS.

\begin{figure*}[t]
\newcommand{\picHeight}{0.37}
\vspace{-4ex} \centering 
\subfloat[]{
\includegraphics[height=0.18\paperwidth, natheight=1804, natwidth=1693]
{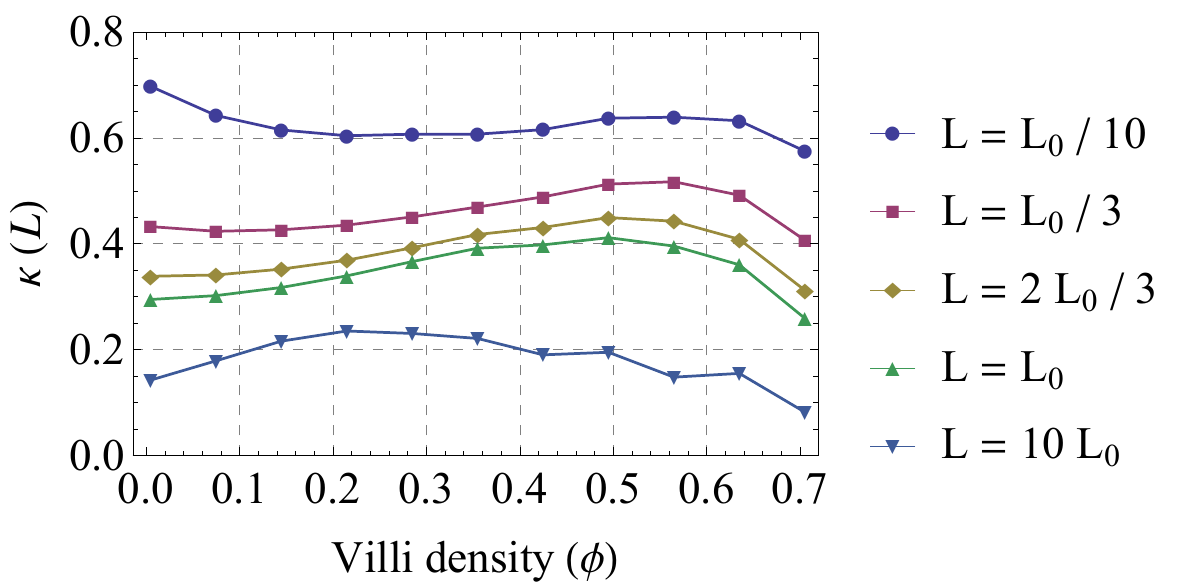}
\label{fig:kappa_L_phi} }
\hspace{4ex}
\subfloat[]{
\includegraphics[height=0.17\paperwidth, natheight=1804, natwidth=1693]
{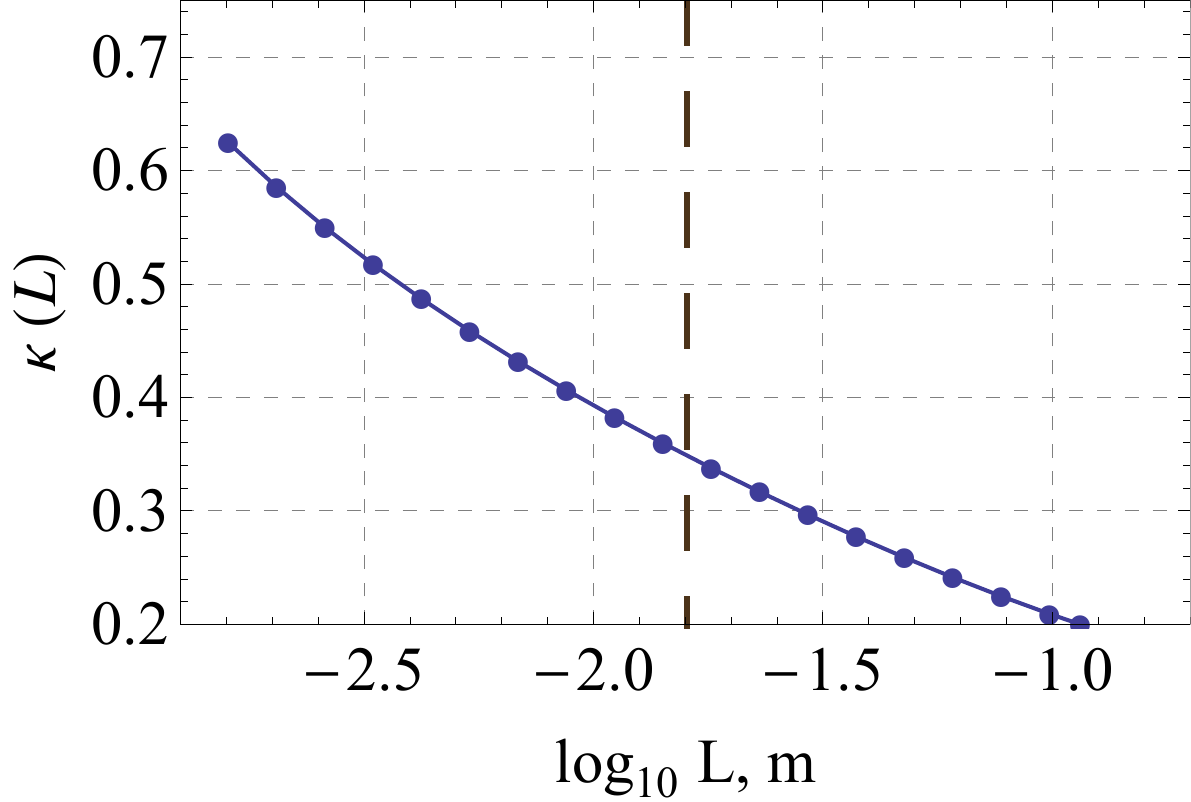}
\label{fig:kappa_L} 
}
\caption{
Dependence of~$\kappa(L)$ on the stream-tube cross-sectional geometry for circular villi in a circular stream tube~(Fig.~\ref{fig:CalculatedGeometries}).
\protectedSubref{fig:kappa_L_phi}: Dependence of~$\kappa(L)$ on villi density~$\phi\equiv\sFet/\sTot$ in a large range of stream-tube lengths. Here~$\sFet$ is the area of the cross-section occupied by fetal villi,~$\sTot$ is the total area of a cross-section, and~$L_0$ is the average stream-tube length~(see Table~\ref{tblCalculationsParameters}). One can see that in the first approximation,~$\kappa(L)$ may be assumed independent of villi configuration. 
\protectedSubref{fig:kappa_L}: Dependence of~$\kappa(L)$ averaged over~$\phi$ on the stream-tube length~$L$. Dashed vertical line denotes the average stream-tube length~$L_0$
}
\end{figure*}

In the first approximation, the coefficient~$w\PAbs/(D\SM)$ in Eq.~\fm{fm:LambdaProperty} describes the dependence of~$\Lambda_j$ on the cross-sectional geometry. Introducing~$q_j$ and the dimensionless length $\ell(L)\equiv \frac {w\PAbs}{uB\SM}L$ into Eq.~\fm{fm:alphaFirstExpression}, we transform it into
\begin{align}
&\alpha
=\frac {w\PAbs}{uB\SM} \kappa(L),
\label{fm:parameter_alpha}
\\
\nonumber
&\text{where\quad}
\kappa(L)\equiv -\frac1{\ell(L)} \ln\left(\sum\limits_{j=1}^\infty \frac{a_j^2}\SM\exp(-\ell(L) q_j^2)\right)
\end{align}
is a dimensionless coefficient depending on~$L$ and the cross-sectional geometry and containing fine details of a given villi distribution and shapes, which are ignored by the integral parameters~$P$ and~$\SM$.
Figure~\ref{fig:kappa_L_phi} shows the dependence of~$\kappa(L)$ on villi density~($\phi\equiv1-\SM/\sTot=\sFet/\sTot$, the part of the cross-section occupied by fetal villi) as calculated numerically for a stream-tube of circular cross-section filled with circular villi. %
One can see that in a first approximation,~$\kappa(L)$ can be considered as independent of the cross-sectional geometry in a wide range of biologically relevant stream-tube lengths. %
For each~$L$, the value of~$\kappa$ can be determined from Fig.~\ref{fig:kappa_L}. 
For the average stream-tube length~($L_0=\unit[1.6]{cm}$, Table~\ref{tblCalculationsParameters}),~$\kappa\approx0.35$. Using Eqs~\fm{fm:parameter_A} and~\fm{fm:parameter_alpha}, the approximate oxygen uptake~\fm{fmApproximateFExpression} can then be rewritten as

\setlength{\mylength}{-1ex}
{
\vspace{\mylength}
\small
\begin{equation}
\label{fm:TheoreticalApproximationWithAAndAlpha}
\FApprox(L)=c_0uB\SM\left(1-\exp\left(-\frac{w PL}{uB\SM}\kappa(L)\right)\right).
\end{equation}
}

\subsection{Dimensionless geometrical parameters}
\label{sect:DimensionlessGeometricalParameters}

In Eq.~\fm{fm:TheoreticalApproximationWithAAndAlpha}, both geometrical parameters~$\lFet$ and~$\SM$ depend on the size of the analyzed region. To facilitate physical analysis of the approximate solution and its comparison to experimental data, we identify two geometrical characteristics of a placental cross-section that are independent of the size of the region:

\begin{itemize}
\item \emph{The fraction of the cross-section occupied by fetal villi}, which we define as the ratio of the total area of villi in a cross-section to the total area of the cross-section: $\phi\equiv\sFet/\sTot$.

\item
\emph{The effective villi radius}, which we define as~$\rEff\equiv2\sFet/P\equiv 2\phi\SM/(P(1-\phi))$. In morphometric studies, the inverse parameter~$2/\rEff=P/\sFet$ is known as \enquote{villi surface density}. For circular villi of radius~$r$,~$\rEff\equiv r$. The mean value of~$\rEff$ for the human placenta can be found in Table~\ref{tblCalculationsParameters}.

\end{itemize}

\begin{figure*}[tb]
\newcommand{\picWidth}{0.37}
\vspace{-4ex} \centering 
\subfloat[]{
\includegraphics[width=\picWidth\paperwidth, natheight=1804, natwidth=1693]
{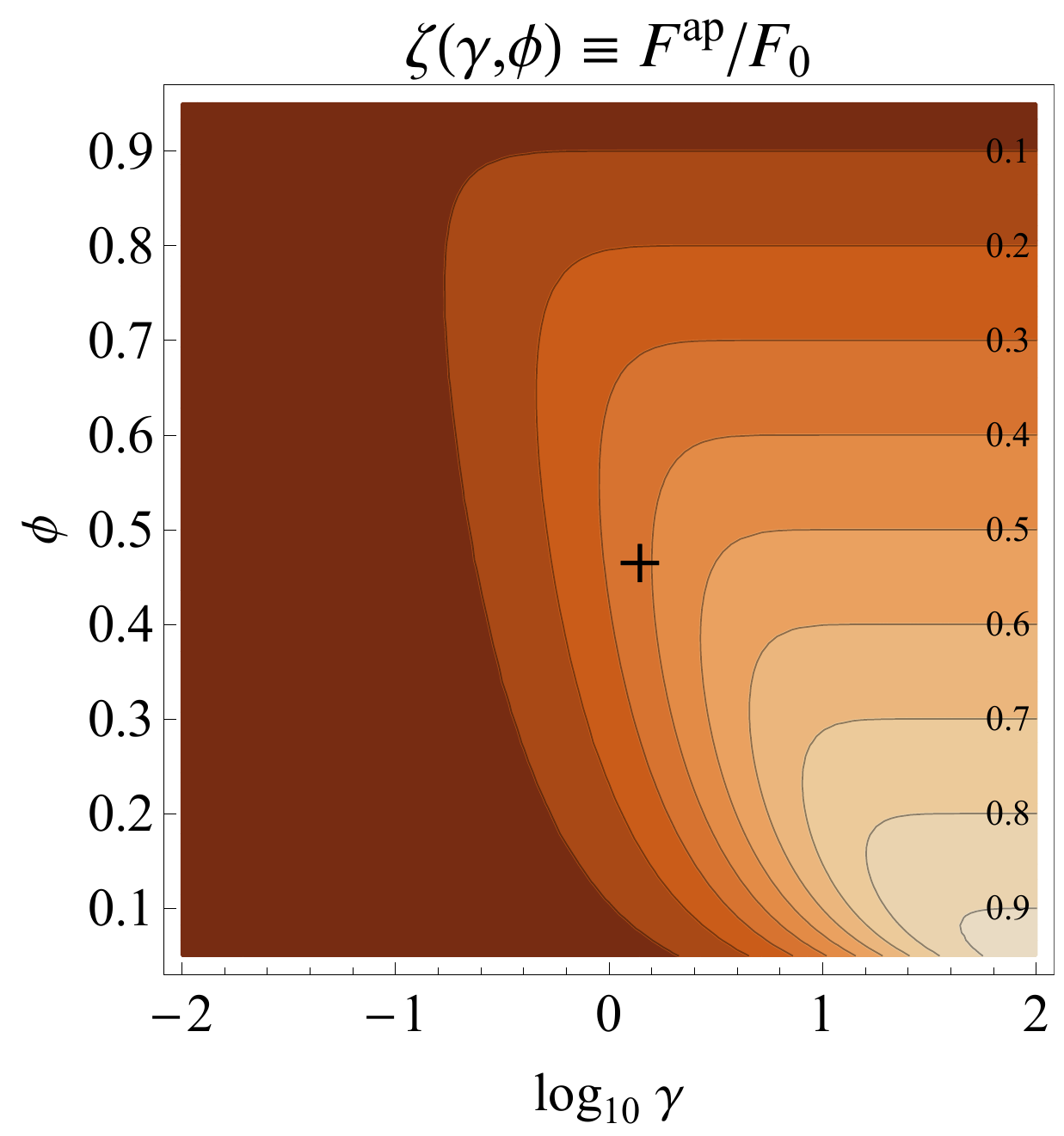}
\label{fig:efficiency_abs} }
\hfill
\subfloat[]{
\includegraphics[width=\picWidth\paperwidth, natheight=1804, natwidth=1693]
{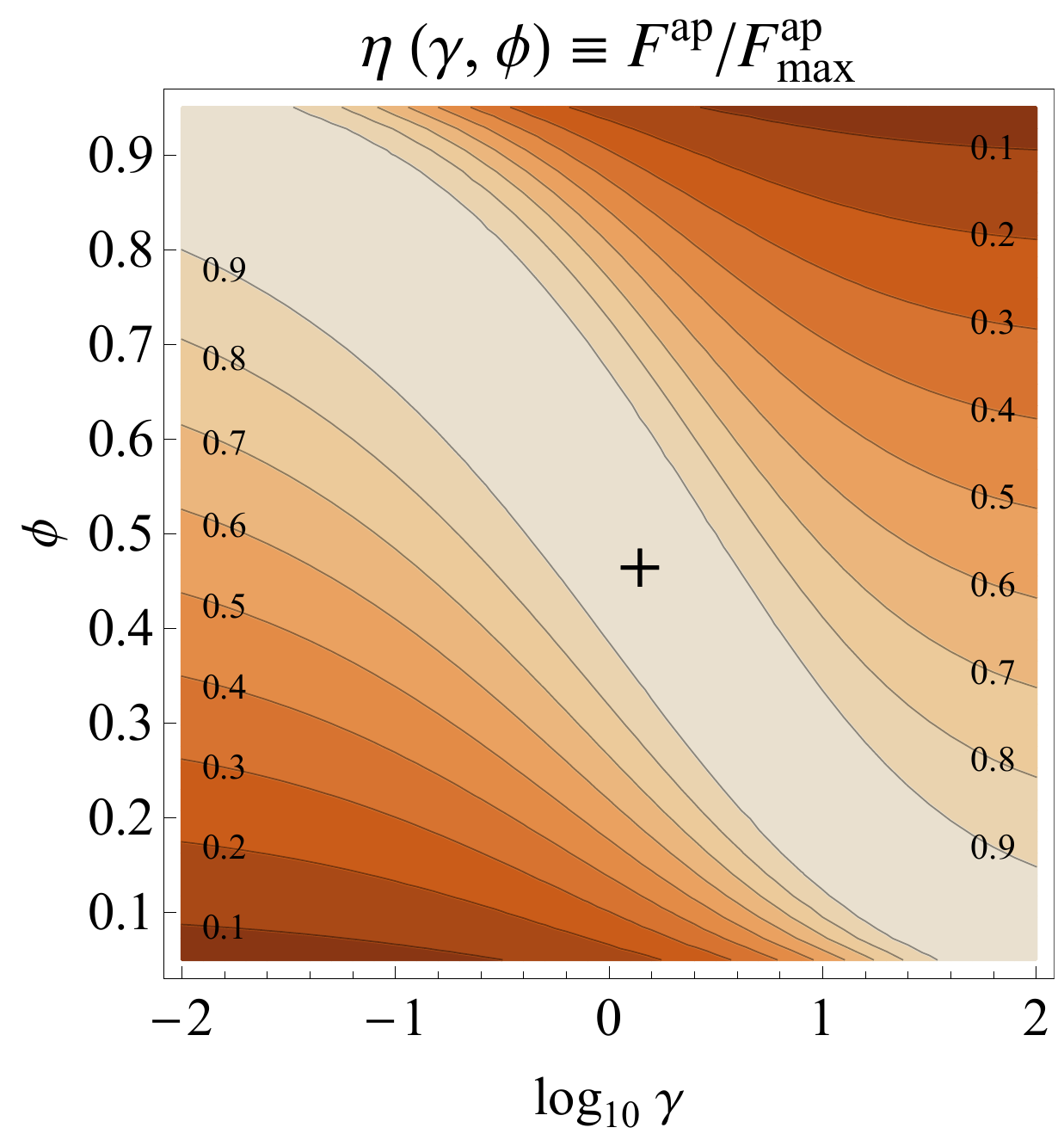}
\label{fig:villi_density_optimality_diagram}
}
\caption{
Diagrams of \emph{oxygen extraction efficiency}~$\zeta(\gamma,\phi)\equiv\FApprox/F_0$~\protectedSubref{fig:efficiency_abs}
and \emph{villi density efficiency}~$\etaE(\gamma,\phi)=\FApprox/\FApproxMax$~\protectedSubref{fig:villi_density_optimality_diagram} as functions of villi density~$\phi$ and the dimensionless parameter~$\gamma$.
The plus symbol marks the parameters $\gamma\approx1.4$~(Sect.~\ref{sect:ParametersCombinations}) and~$\phi\approx0.46$~\citep{Serov2015Optimality} expected on the average in a healthy human placenta~($\kappa=0.35$)
}
\label{fig:efficiency_diagrams}
\end{figure*}

Substitution of these definitions into Eq.~\fm{fm:TheoreticalApproximationWithAAndAlpha} gives
\begin{align}
\zeta(\gamma,\phi)&\equiv\frac{\FApprox(\gamma,\phi)}{F_0}
\nonumber
\\
&=(1-\phi)\left(1-\exp\left(-\gamma(\rEff,L)\frac{\phi}{1-\phi}\right)\right),
\label{fm:FinalTheoreticalApproximationEpsilon}
\end{align}
\begin{flalign}
&\text{where\ }
&&\gamma(\rEff,L)
\equiv\frac{2w\kappa}{uB\rEff}L,
&&F_0\equiv c_0uB\sTot.
&&
\label{fm:GammaAndF0Definitions}
\end{flalign}
The physical meaning of~$F_0$ follows from its definition: it is the maximal oxygen flow entering a stream tube that would be achieved if no villi obstructed the~IVS of the stream-tube. %
The incoming flow in the presence of villi is~$\FInfty\equiv F_0(1-\phi)$. 
Note that~$\gamma$ includes information on the average villus shape through the parameter~$\rEff$, and that~$F_0$ is independent of the cross-sectional geometry.
Normalized oxygen uptake $\zeta(\gamma,\phi)\equiv\FApprox/F_0$ (which we will call \emph{oxygen extraction efficiency}) is plotted in Fig.~\ref{fig:efficiency_abs}.
The physical meaning of~$\gamma$ is discussed later.

\subsection{Optimal cross-sectional geometry}

Looking at Fig.~\ref{fig:efficiency_abs} and expression~\fm{fm:FinalTheoreticalApproximationEpsilon} for oxygen uptake, it is natural to ask
what are the \enquote{optimal} values of the parameters~$\phi$ and~$\gamma$ that maximize oxygen uptake at a given stream-tube length. 
Figure~\ref{fig:efficiency_abs} clearly shows two trends:
\begin{itemize}
\item  
for any fixed value of~$\phi$, larger~$\gamma$ provides larger uptake;

\item
for any fixed value of~$\gamma$, there exists some intermediate value of~$\phi$: $0<\phiOpt(\gamma)<1$ that maximizes oxygen uptake. This optimal value~$\phiOpt(\gamma)$ tends to diminish with~$\gamma$.
\end{itemize}
Note that from the definition~\fm{fm:GammaAndF0Definitions} it follows that in terms of cross-sectional geometrical parameters, an increase of~$\gamma$ corresponds to a decrease of the effective radius~$\rEff$. 
Figure~\ref{fig:efficiency_abs} can then be interpreted in the following way: \emph{it is more efficient to have many small villi than fewer big villi occupying the same area}. This prediction can be understood if one considers the fact that small villi have more absorbing surface per unit of cross-sectional area.
 
However, in the human placenta,~$\rEff$ cannot be infinitely small~(and hence~$\gamma$ infinitely large). Indeed, villi possess an internal structure~(e.g., fetal blood vessels) to transport the absorbed oxygen to the fetus. The decrease of~$\rEff$ below some value is likely making the villi less efficient in transporting the already absorbed oxygen. 
This argument is supported by an experimental observation that in the terminal and mature intermediate villi of the human placenta~(the smallest villi), blood vessels normally occupy the main part of the internal volume. %
The mean radius of these smallest villi is~$r\approx\unit[25\text{--}30]{\mu m}$~\citep[see Table~28.7 in][]{benirschke_pathology} and is not reported to significantly vary within the same placenta or between different placentas~\citep{benirschke_pathology}. %
At the same time, villi density may exhibit significant spatial fluctuations within the same placenta as well as between different placentas~\citep{Bacon1986}.

It is then reasonable to reformulate the initial question of optimization of the cross-sectional geometry as which villi density provides the highest oxygen uptake for a given effective villi radius~$\rEff$. Mathematically, it is the question of finding the maximum of~$F$ against~$\phi$ for a fixed~$\gamma(\rEff)$.

\subsection{Optimal villi density}

Under the constraint of fixed~$\gamma$, Eq.~\fm{fm:FinalTheoreticalApproximationEpsilon} implies the existence of maximal uptake at a certain villi density. The reasoning is the following:
\begin{itemize}

\item
$F(\phi=0)=0$, because the condition~$\phi=0$ means no feto-maternal interface and hence no uptake.

\item
$F(\phi=1)=0$, because fetal vessels occupy the entire cross-section of the stream tube and the incoming~MBF is zero as it does not have space to flow.

\item
$F>0$ for $0<\phi<1$, which corresponds to the fact that the placenta transfers oxygen from mother to fetus for intermediate villi densities. Hence, there always exists a maximal oxygen uptake~$\FMax(L)$ at a certain villi density $0<\phiOpt(L)<1$ for any~$\gamma(\rEff)$.

\end{itemize}
Here we have used~$F$ and not~$\FApprox$ symbol for oxygen uptake to underline that these arguments are general and are valid not only for the approximate flow, but for the exact flow as well.

The optimal villi density can then be obtained by solving the equation~%
$\left.\partialFrac{\FApprox(\phi,L)}{\phi}\right|_{\phi=\phiOpt}=0$,
\begin{flalign}
&\text{or\quad}
&&\exp\left(\gamma\frac{\phiOpt}{1-\phiOpt}\right)=1+\frac{\gamma}{1-\phiOpt}.
&&
\label{fmAnalyticProblemWithParametersCombinationDifferentiated}
\end{flalign}

One can note that Eq.~\fm{fmAnalyticProblemWithParametersCombinationDifferentiated} is an explicit equation for~$\phiOpt$ as a function of~$\gamma$ only. As a consequence, it does not require any eigenvalues calculation.

\begin{figure*}[tb]
\newcommand{\picWidth}{0.37}
\vspace{-4ex} \centering 
\subfloat[]{
\includegraphics[width=\picWidth\paperwidth, natheight=1804, natwidth=1693]
{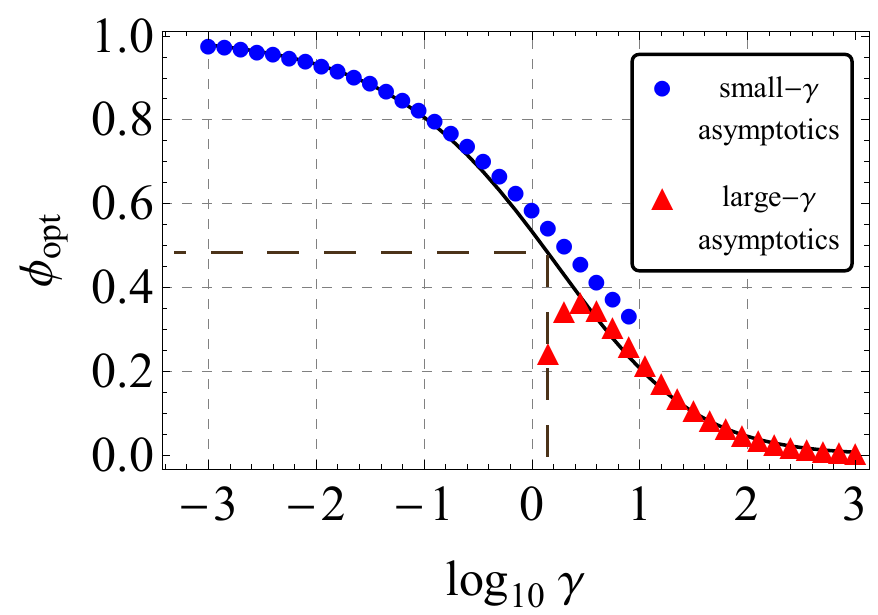}
\label{fig:PhiOptOfGamma} }
\hspace{1ex}
\subfloat[]{
\includegraphics[width=\picWidth\paperwidth, natheight=1804, natwidth=1693]
{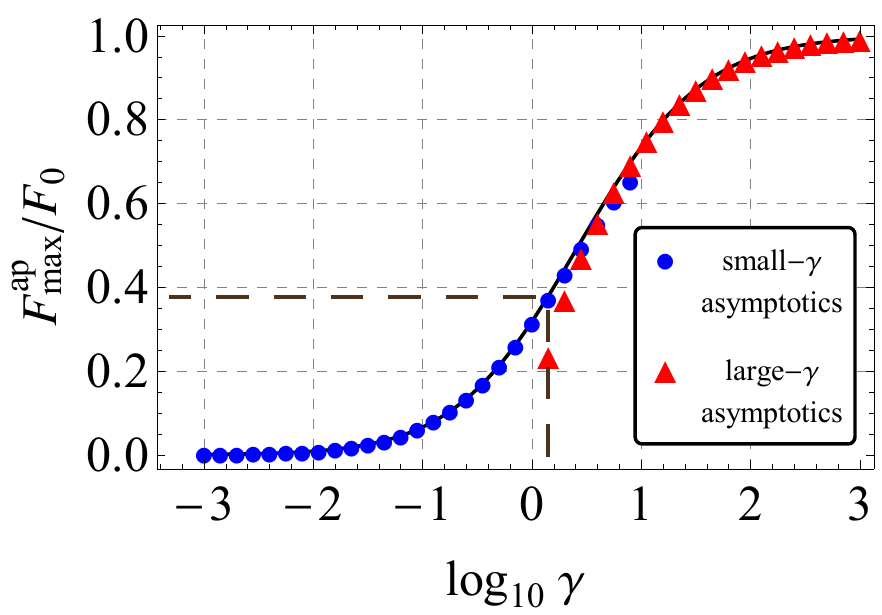}
\label{fig:FMaxOfGamma} }

\caption{Analytical predictions of the optimal villi density~\protectedSubref{fig:PhiOptOfGamma} and of the normalized maximal uptake~\protectedSubref{fig:FMaxOfGamma} as functions of~$\gamma$ (solid lines).
Small~$\gamma$ asymptotics are shown by circles; large~$\gamma$ asymptotics are shown by triangles. 
Dashed lines marks~$\gamma\approx1.4$ observed in a healthy human placenta~(see Sect.~\ref{sect:ParametersCombinations}).
}
\label{fig:AnalyticalResultsDependenceOnGamma}
\end{figure*}

The substitution~$x\equiv \gamma \phiOpt/ (1-\phiOpt)$ reduces Eq.~\fm{fmAnalyticProblemWithParametersCombinationDifferentiated} to the form~%
$\gamma=g(x)$, where\\*$g(x)\equiv\e^x-x-1$, %
and its solution can be represented as~$x=g^{-1}(\gamma)$. %
Although the inverse function~$g^{-1}(\gamma)$ does not have an explicit representation, its form can be easily calculated once and then the tabulated values can be used in practice.
Returning to the definition of~$x$, one obtains~$\phiOpt$ as a function of~$\gamma$:
\begin{equation}
\phiOpt(\gamma)=\frac1{1+{\displaystyle\frac\gamma{g^{-1}(\gamma)}}}.
\label{fm:PhiOptOfGamma}
\end{equation} 
The function~$\phiOpt(\gamma)$~(Fig.~\ref{fig:PhiOptOfGamma}) can be used to calculate an optimal villi density for a placental region if~$\gamma$ is known for this region. Substitution of the last result into Eq.~\fm{fm:FinalTheoreticalApproximationEpsilon} gives the corresponding maximal uptake:
\begin{equation}
\label{fm:FMaxOfGamma}
\frac{\FApproxMax(\gamma)}{F_0}=\gamma\frac{1-\exp(-g^{-1}(\gamma))}{\gamma+g^{-1}(\gamma)}.
\end{equation}

From the asymptotic behavior of~$g(x)$ at small and large~$x$, we obtain the following asymptotic formulas for~$\phiOpt$ and~$\FApproxMax/F_0$:
\begin{align}
&\phiOpt(\gamma)\simeq
\left\{
\begin{aligned}
&\frac1{1+\sqrt{\gamma/2}}, &&\gamma\ll1
\\
&\ln(\gamma) / \gamma, &&\gamma\gg1
\end{aligned}
\right.
\\
&\frac{\FApproxMax(\gamma)}{F_0}\simeq
\left\{
\begin{aligned}
&\frac{1-\e^{-\sqrt{2\gamma}}}{1+\sqrt{2/\gamma}}, &&\gamma\ll1
\\
&\frac{1-1/\gamma}{1+\ln(\gamma)/\gamma}, &&\gamma\gg1
\end{aligned}
\right.
\label{fm:Asymptotics}
\end{align}
Figure~\ref{fig:AnalyticalResultsDependenceOnGamma} shows that these asymptotics accurately approximate~$\phiOpt(\gamma)$ and~$\FApproxMax(\gamma)/F_0$ not only in the limits of~$\gamma\ll1$ and~$\gamma\gg1$, but for all~$\gamma$. For instance, it can be calculated that if \mbox{small-$\gamma$} asymptotic is used for~$\gamma\leqslant3$ and \mbox{large-$\gamma$} asymptotic is used for~$\gamma>3$, the maximal relative error of the second formula of Eq.~\fm{fm:Asymptotics} is less than~\unit[4]{\%}. 
In other words, we obtained simple explicit approximations for the optimal villi density~$\phiOpt$ and the normalized maximal oxygen uptake~$\FApproxMax/F_0$ as functions of a single parameter~$\gamma$.

\subsection{Villi density efficiency}

Basing on the optimal villi density and the maximal uptake introduced in the previous section, one can define a quantitative measure of optimality of villi density in a given~(not optimal) geometry.
 
If the given geometry is characterized by the parameters~$(\gamma,\phi)$, its \emph{villi density efficiency} can be defined as the ratio of oxygen uptake in this particular geometry to the maximal value, which can be obtained with the same~$\gamma$~(Fig.~\ref{fig:FMaxOfGamma}):
\begin{align}
\etaE(\gamma,\phi)&\equiv\frac{\FApprox(\gamma,\phi)}{\FApproxMax(\gamma)}
\nonumber
\\
&=\frac{(1-\phi)\left(1-\exp\left(-\gamma\frac\phi{1-\phi}\right)\right)}{(1-\phiOpt(\gamma))\left(1-\exp\left(-\gamma\frac{\phiOpt(\gamma)}{1-\phiOpt(\gamma)}\right)\right)}.
\label{fm:eta_definition}
\end{align}
Figure~\ref{fig:villi_density_optimality_diagram} presents the villi density efficiency~$\eta(\gamma,\phi)$ in a physiological range of~$(\gamma,\phi)$ and at~$L=L_0$. 

Following the comment after Eq.~\fm{fm:Asymptotics}, Eq.~\fm{fm:eta_definition} can be rewritten as

\setlength{\mylength}{-1ex}
{
\vspace{\mylength}
\footnotesize
\begin{equation*}
\etaE(\gamma,\phi)\simeq\left\{
\begin{aligned}
\frac{(1-\phi)\left(1-\exp\left(-\gamma\frac\phi{1-\phi}\right)\right)(1+\sqrt{2/\gamma}}{1-\e^{-\sqrt{2\gamma}}}, &&\gamma\leqslant3
\\
\frac{(1-\phi)\left(1-\exp\left(-\gamma\frac\phi{1-\phi}\right)\right)(1+\ln(\gamma)/\gamma)}{1-1/\gamma}, &&\gamma>3
\end{aligned}
\right.
\end{equation*}
\vspace{\mylength}
}

with a maximal relative error of~\unit[4]{\%}. Note that the last equation does not require calculation of~$\phiOpt$ and is an explicit function of~$\gamma$ and~$\phi$.

Note finally that the optimality indicators~$\zeta$ and~$\etaE$ play different roles. 
Oxygen extraction efficiency~$\zeta$~(Fig.~\ref{fig:efficiency_abs}) indicates the fraction of the maximal possible incoming oxygen flow~$F_0$ that is absorbed by a given cross-sectional geometry. 
The higher is the value of~$\zeta$, the higher is the \emph{absolute} value of fetal oxygen uptake.
At the same time, villi density efficiency~$\etaE$~(Fig.~\ref{fig:villi_density_optimality_diagram}) shows how far the villi density of a given cross-section is from its optimal value for a fixed~$\gamma(\rEff)$. 
The higher is the value of~$\etaE$, the closer is fetal oxygen uptake to the maximal value for the given villi radius~$\rEff$.


\section{Results}

\begin{figure*}[tbp]
\newcommand{\picWidth}{0.37}  
\vspace{-4ex} \centering 
\subfloat[]{
\includegraphics[width=\picWidth\paperwidth, natheight=477, natwidth=912]
{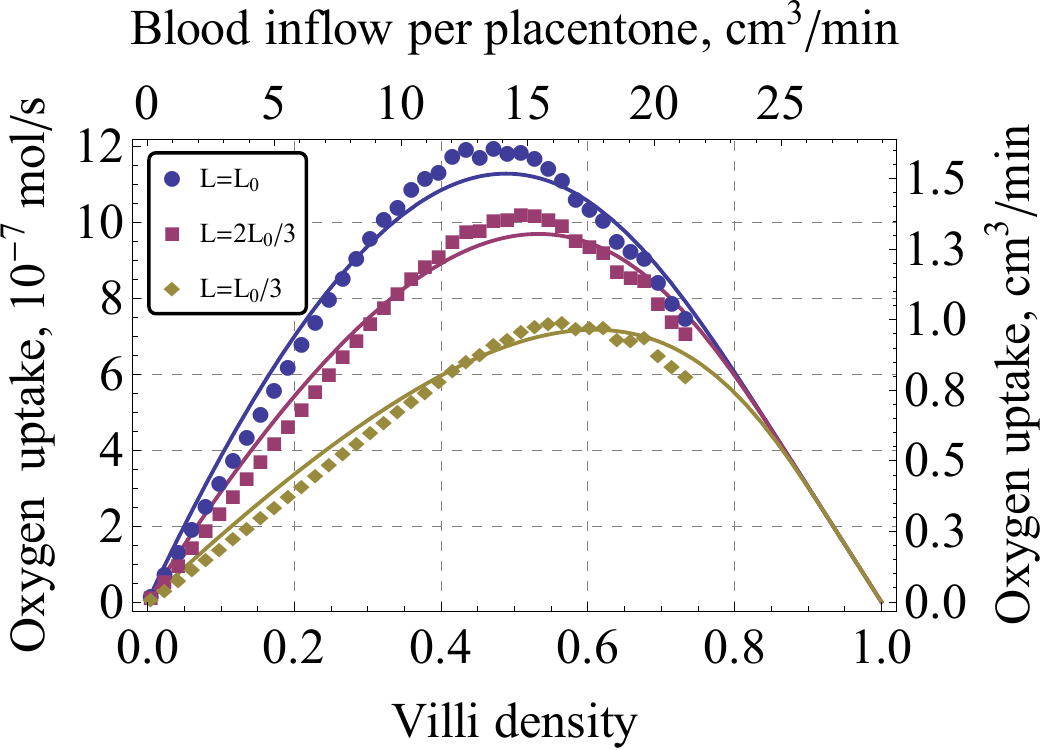}
\label{fig:UptakeVilliDensityFixedL} }
\hfill
\subfloat[]{
\includegraphics[width=\picWidth\paperwidth, natheight=477, natwidth=912]
{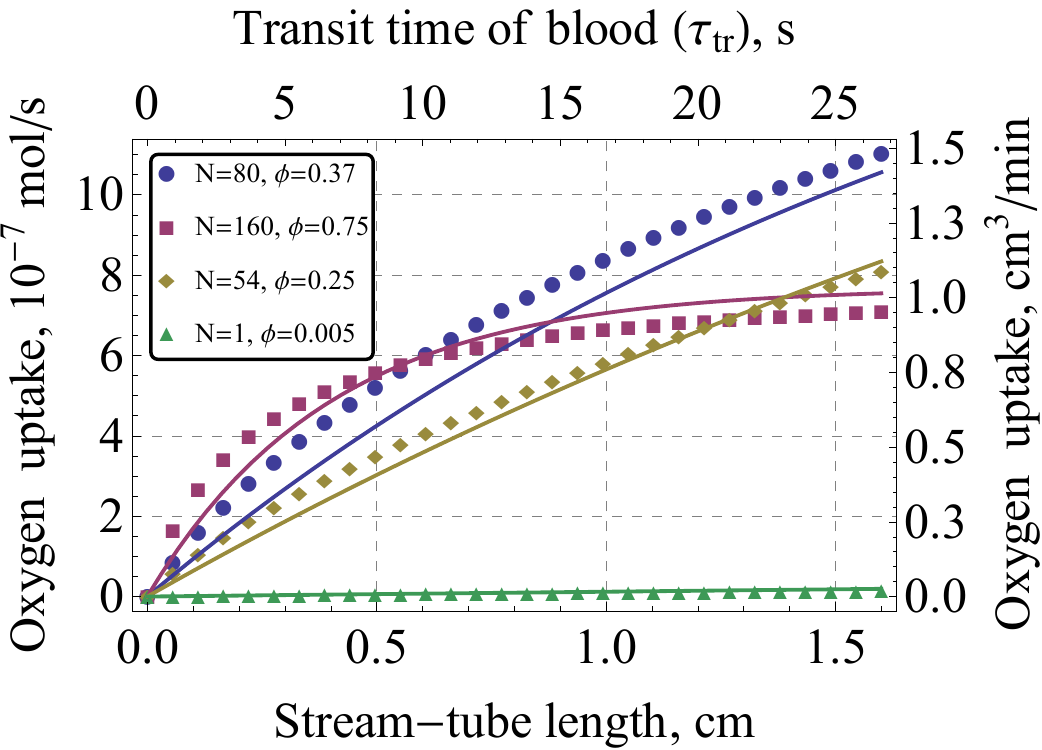}
\label{fig:UptakeLVilliDensity} }
\caption{Oxygen uptake of a single placentone as a function of geometrical parameters. Solid lines correspond to the analytical approximation; symbols reproduce the results of numerical simulations of~\citet{Serov2015Optimality}.
\protectedSubref{fig:UptakeVilliDensityFixedL}
Oxygen uptake as a function of villi density~$\phi$ for three lengths~$L$: $L_0/3$, $2L_0/3$, and $L_0$. For each of these lengths, the corresponding value of~$\kappa$ was determined from Fig.~\ref{fig:kappa_L}:~$\kappa\approx\{0.46,0.39,0.35\}$ for~$L=\{L_0/3,2L_0/3,L_0\}$ respectively. 
MBF velocity $u=\unit[0.6]{mm/s}$ was used.
It can be observed that peak uptake moves to smaller villi densities for larger stream-tube lengths~$L$. 
Note that in the analytical theory oxygen uptake is calculated directly for the radius~$R$, whereas in the numerical simulation oxygen uptake is calculated for~$\RNum$ and then rescaled to the placentone radius~$R$ by multiplying by~$R^2/\RNumSquared$. %
Note also that the numerical curves do not go beyond the villi density~$\phi\approx0.75$ because there exists a maximal packing density of circles in a large circle, and numerical results cannot be calculated beyond that density~\citep[see][]{Specht2009}. The analytical theory, on the contrary, does not rely on particular shapes or distributions of villi, but operates only with villi density and the effective villi radius, thus allowing the results to be calculated for villi densities beyond this limit (although in the region of~$\phi>0.75$ villi cannot be circular, the same~$\rEff$ is maintained). 
\protectedSubref{fig:UptakeLVilliDensity}
Oxygen uptake as a function of stream-tube length~$L$ for a fixed villi density~$\phi$. 
Small deviations of the theory from the numerical results seen in the figure are explained by the fixed~$\kappa=\kappa(L_0)\approx0.35$ used for all lengths.
Various symbols represent villi densities of Fig.~\ref{fig:CalculatedGeometries}
}
\label{fig:UptakeResults}
\end{figure*}

\begin{figure*}[t]
\center{\includegraphics[width=0.5\paperwidth, natheight=477, natwidth=912]
{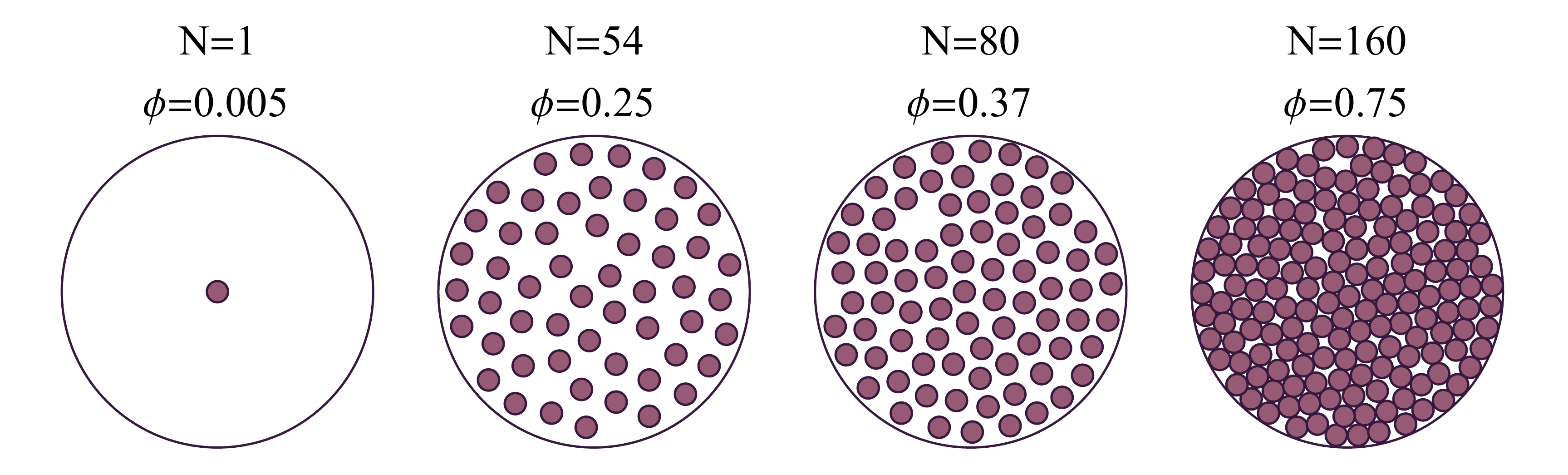}}
\caption{Villi distributions for which oxygen uptake was calculated in~\citet{Serov2015Optimality}. Number of villi~($N$) and the corresponding villi density~($\phi$) are displayed above each case. The analytical theory was applied to the corresponding~$\phi$ with~$\rEff$ given in Table~\ref{tblCalculationsParameters}. Maternal blood flows in the white space in the direction perpendicular to the cross-sections.
}
\label{fig:CalculatedGeometries}
\end{figure*}

\begin{figure*}[tb]
\newcommand{\picWidth}{0.37}
\vspace{-4ex} \centering 
\subfloat[]{
\includegraphics[width=\picWidth\paperwidth, natheight=1804, natwidth=1693]
{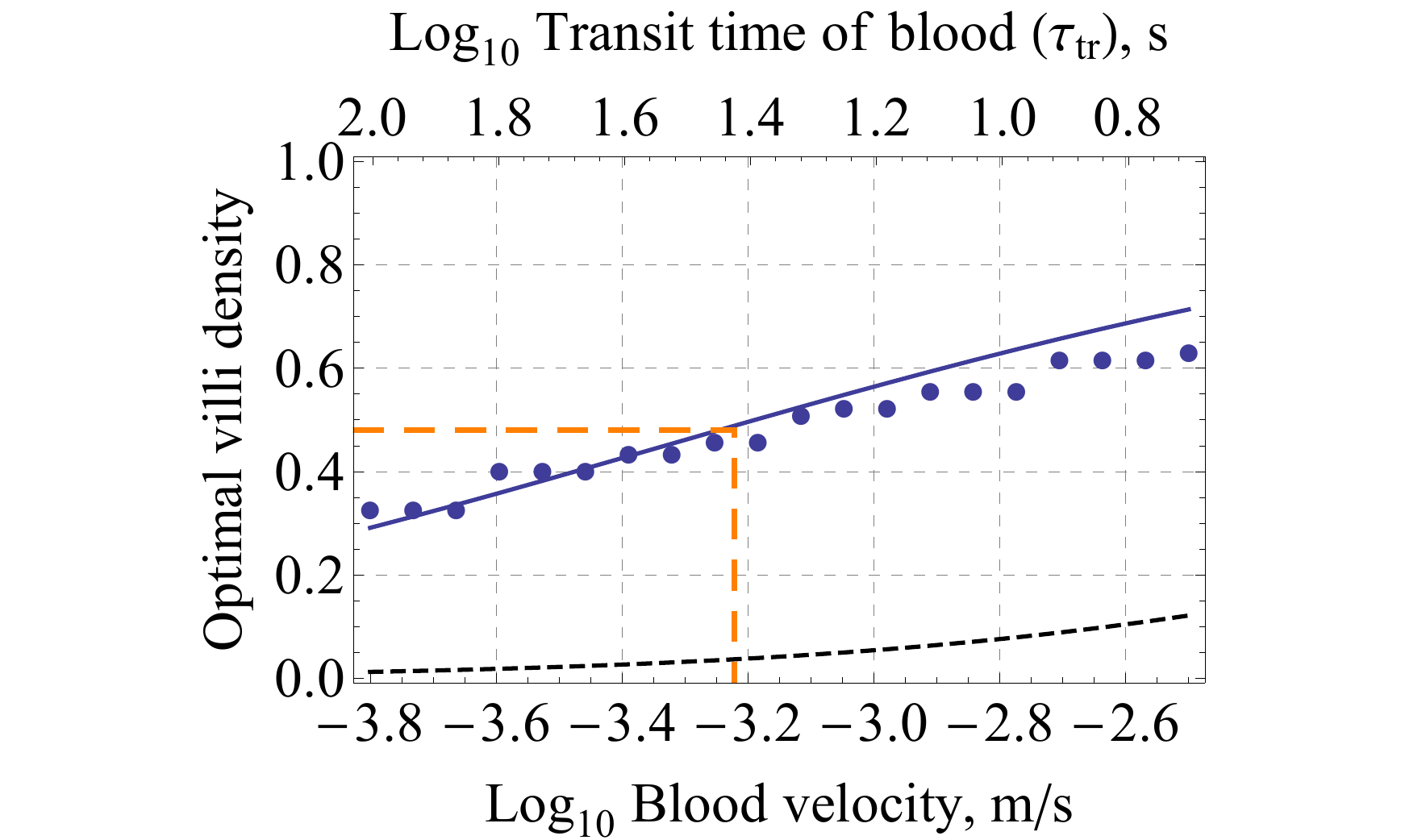}
\label{fig:VilliDensityVelocity} }
\hspace{1ex}
\subfloat[]{
\includegraphics[width=\picWidth\paperwidth, natheight=1804, natwidth=1693]
{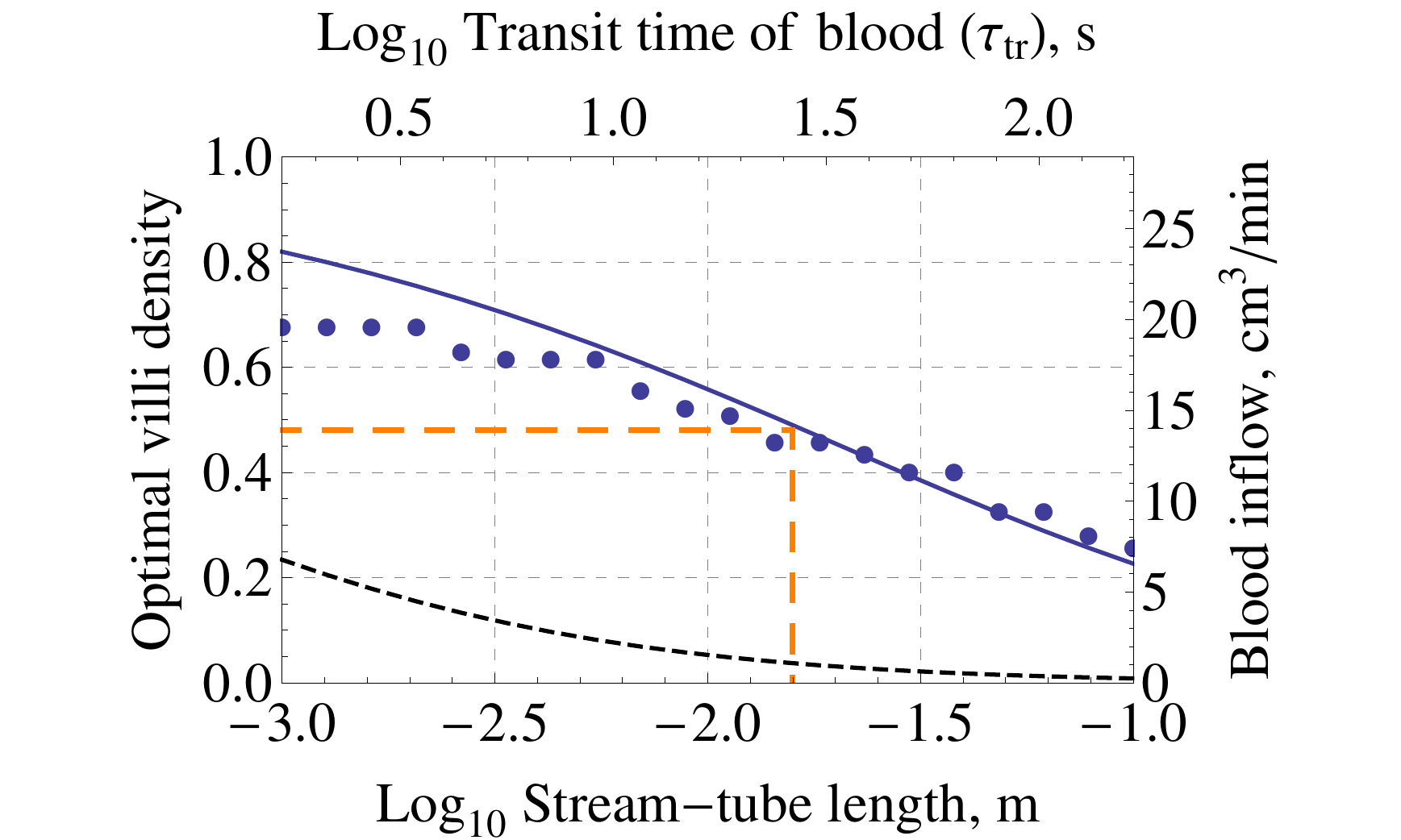}
\label{fig:VilliDensitySystemLength} }
\vspace{-2ex}
\subfloat[]{
\includegraphics[width=\picWidth\paperwidth, natheight=1804, natwidth=1693]
{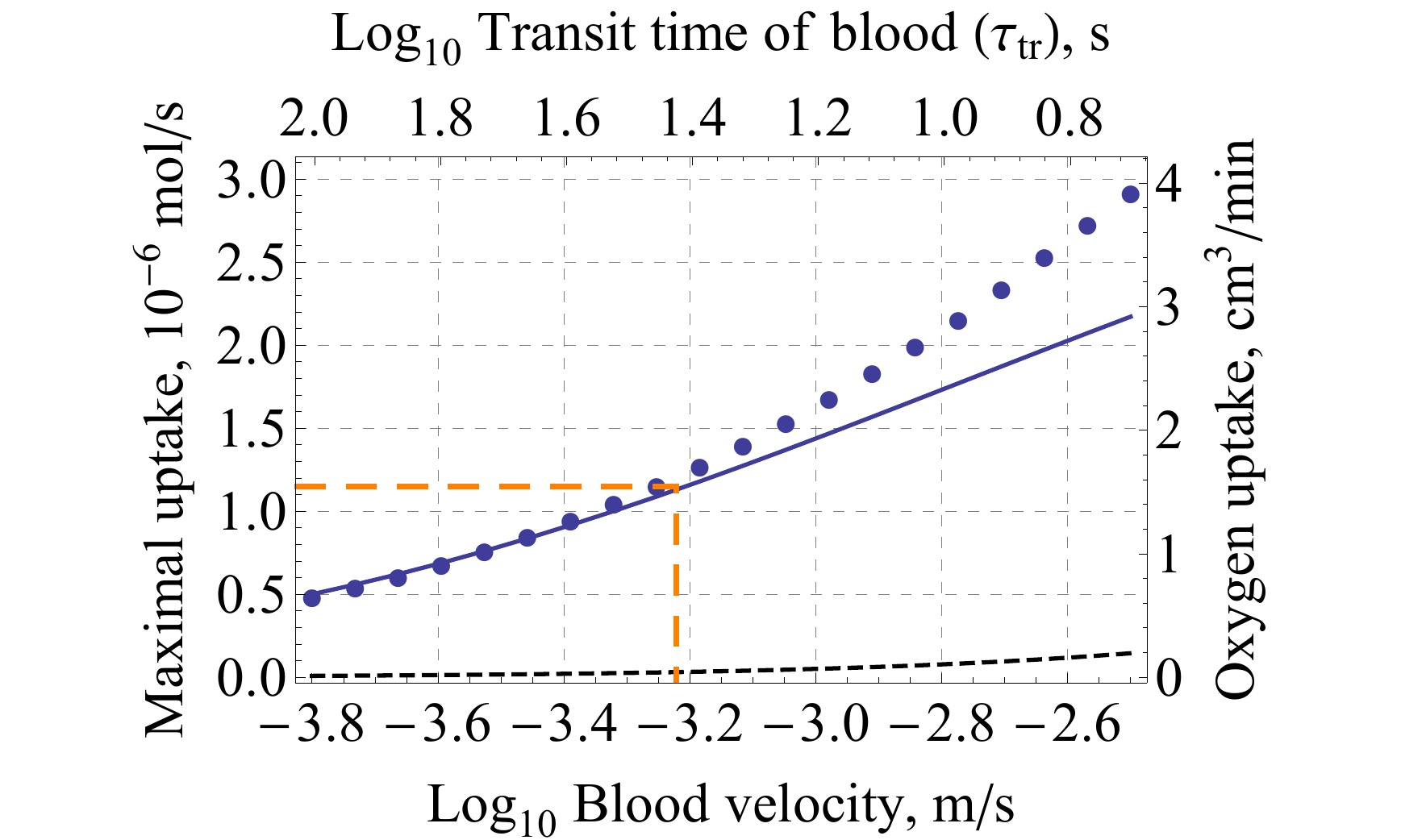}
\label{fig:MaximalUptakeVelocity} }
\hspace{1ex}
\subfloat[]{
\includegraphics[width=\picWidth\paperwidth, natheight=1804, natwidth=1693]
{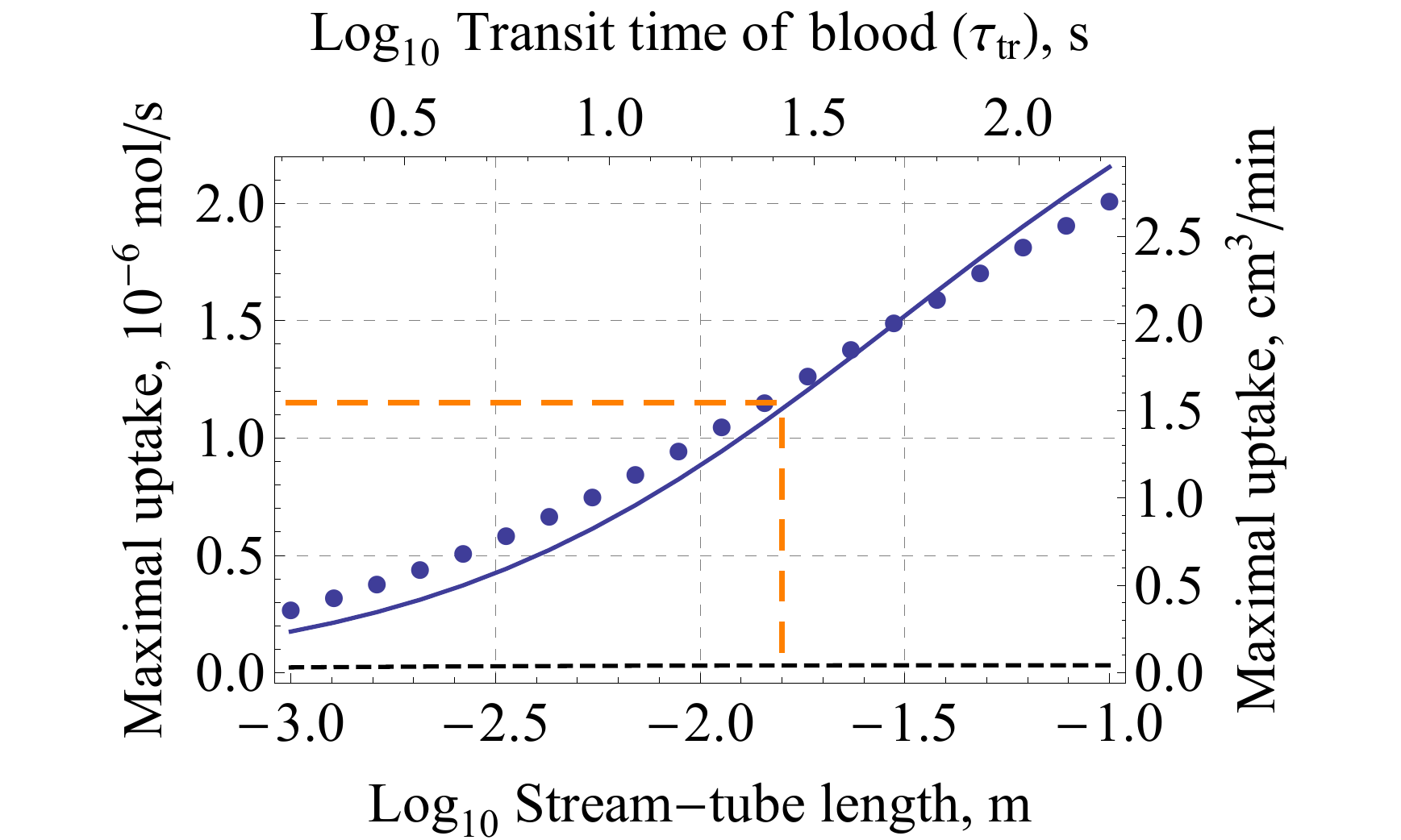}
\label{fig:MaximalUptakeSystemLength} }

\caption{
\protectedSubref{fig:VilliDensityVelocity}, \protectedSubref{fig:MaximalUptakeVelocity}:
Dependence of the optimal villi density~\protectedSubref{fig:VilliDensityVelocity} and the maximal oxygen uptake~\protectedSubref{fig:MaximalUptakeVelocity}
on the MBF velocity at a fixed length~$L_0$~(Table~\ref{tblCalculationsParameters}) for a single placentone. 
\protectedSubref{fig:VilliDensitySystemLength},~\protectedSubref{fig:MaximalUptakeSystemLength}:
Dependence of the optimal villi density~\protectedSubref{fig:VilliDensitySystemLength} and the maximal oxygen uptake~\protectedSubref{fig:MaximalUptakeSystemLength}
on stream-tube length at a fixed~MBF velocity~$u$~(Table~\ref{tblCalculationsParameters}) for a single placentone. 
Solid curves represent analytical results, while filled circles correspond to numerical simulations for circular villi~(Fig.~\ref{fig:CalculatedGeometries}).
Dashed curves show analytical results in blood with no hemoglobin~($B=1$, as in artificial perfusion experiments).
Straight dashed lines indicate the expected average~MBF velocity~$u$ and stream-tube length~$L_0$~(Table~\ref{tblCalculationsParameters}). 
Step growth of numerical results seen in Figs~\protectedSubref{fig:VilliDensityVelocity}, \protectedSubref{fig:VilliDensitySystemLength} is explained by discrete changes of villi density in the numerical simulations due to discrete changes in the number of villi in the cross-section
}
\label{fig:ResultsDependenceOnVelocity}
\end{figure*}

Figure~\ref{fig:UptakeResults} shows that fetal oxygen uptake predicted by the analytical Eq.~\fm{fm:FinalTheoreticalApproximationEpsilon} agrees well with numerically calculated results~\citep{Serov2015Optimality} in wide ranges of stream-tube lengths~($L$) and villi densities~($\phi$). Figure~\ref{fig:UptakeVilliDensityFixedL} demonstrates the existence of maximal oxygen uptake corresponding to an optimal villi density for each stream-tube length. 
The value of~$\kappa(L)$ was determined from Fig.~\ref{fig:kappa_L} for each considered length~$L$. 
These results were calculated for the same geometries as in our earlier numerical simulation~(Fig.~\ref{fig:CalculatedGeometries}). 
We emphasize that numerical simulations with identical circular villi are shown only for the purpose of validation. 
The proposed analytical theory, which uses only villi density and the effective villi radius~($\rEff$) as geometrical information, is applicable to villi of arbitrary shapes and sizes.
The agreement of analytical curves with numerical points shows that for uniform villi distributions, knowing villi density and the effective villi radius is enough to predict the oxygen uptake.

Variations of the parameter~$\gamma$ in analytical expressions for optimal villi density and maximal uptake~(Eqs~\fm{fm:PhiOptOfGamma},~\fm{fm:FMaxOfGamma}) can be interpreted in terms of changes of individual parameters of the model, other parameters being fixed. For example, optimal villi density and maximal uptake can be plotted as functions of~MBF velocity~(Figs~\ref{fig:VilliDensityVelocity},~\ref{fig:MaximalUptakeVelocity}) or stream-tube length~(Figs~\ref{fig:VilliDensitySystemLength} and~\ref{fig:MaximalUptakeSystemLength}). %
An agreement between the plotted curves and numerical results of~\citet{Serov2015Optimality} can be observed.

All four plots in Fig.~\ref{fig:ResultsDependenceOnVelocity} feature a dashed black curve representing a fictitious case of blood having no hemoglobin but transporting only oxygen dissolved in the blood plasma. Mathematically, this case is described by oxygen-hemoglobin dissociation parameter~$B=1$, which is about~100 times smaller than that for blood with~Hb. As predicted by Eq.~\fm{fm:PhiOptOfGamma}, the no-Hb curves for optimal villi density have the same shape but are shifted by two orders of magnitude as compared to those for normal blood.


\section{Discussion}

\subsection{Parameters~$\gamma$ and~$F_0$}
\label{sect:ParametersGammaAndF0}

\subsubsection{Values}

Taking~$\pi R^2$ as the total area of the cross-section, parameters from Table~\ref{tblCalculationsParameters} and~$\kappa(L_0)\approx0.35$ for the average stream-tube length~$L_0\approx\unit[1.6]{cm}$~(Fig.~\ref{fig:kappa_L}), from Eq.~\fm{fm:GammaAndF0Definitions} one can estimate the values of~$\gamma$ and~$F_0$ %
which characterize a \enquote{healthy} regime of our placenta model:~$\gamma\approx1.4$, $F_0\approx\unit[3\cdot10^{-6}]{mol/s}$. %
\label{sect:ParametersCombinations}%
The obtained average value of~$\gamma$ together with the average villi density~$\phi\approx0.46$~\citep{Serov2015Optimality} are marked by crosses in the diagrams in Fig.~\ref{fig:efficiency_diagrams}. 
One can see that the theory predicts that an average placenta extracts around~\unit[35]{\%} of the maximal possible incoming oxygen flow~$F_0$~(Fig.~\ref{fig:efficiency_abs}), and that this value is close to the maximal one for the given effective villi radius~(Fig.~\ref{fig:villi_density_optimality_diagram}). 
Although~\unit[35]{\%} seems to be a low value, note that oxygen extraction efficiency of~\unit[100]{\%} is never achievable since in the presence of villi only a part of the flow unobstructed by villi~($F_0$) can be transferred to the fetus.

To have predictive power, $\gamma$ and~$\phi$ need to be measured for different healthy as well as pathological placentas over the whole exchange region. 
Such measurements require development of image analysis techniques, which could automatically determine these characteristics for histological placental slides. %
Such measurements have not yet been performed and present an important perspective to this study. 
At the same time, because of the lack of experimental information about several other parameters~(namely,~$u$, $w$, $L$) in each studied placenta, correlations of changes of~$\gamma$ and~$\phi$ with changes of fetal development characteristics~(such as birth-weight, placenta weight or their ratio) are expected to be of more practical use than absolute values of~$\gamma$ and~$\phi$. 
Note finally that the optimal geometry and maximal uptake may change for non-slip boundary conditions; further studies are required to clarify this point~\citep[for discussion see][]{Serov2015Optimality}.

\subsubsection{Parameter~$\gamma$}

The two parameters~$\gamma$ and~$F_0$ play different roles. According to Eq.~\fm{fmAnalyticProblemWithParametersCombinationDifferentiated},~$\gamma$ alone determines the optimal villi density, while~$F_0$ together with~$\gamma$ determines the maximal oxygen uptake~(Eq.~\fm{fm:FMaxOfGamma}). %
 
A clear physical interpretation of~$\gamma$ can be obtained by rewriting~\fm{fm:GammaAndF0Definitions} as
\begin{equation*}
\gamma=\frac{L/u}{B\rEff/(2w\kappa)}
=\frac{\tauTransit}{\tauEff},
\end{equation*}
where~$\tauTransit\equiv L/u$ is the transit time of maternal blood through the placenta~(while it flows along a stream tube of length~$L$ with an average velocity~$u$) and~$\tauEff\equiv B\rEff/(2w\kappa)$ is oxygen extraction time of a placental cross-section. As a consequence,~$\gamma$ can be understood as a quantitative measure of balance between two oxygen transport mechanisms: the longitudinal convective flow and the transverse diffusion. 
In other words,~$\gamma$ \emph{describes the level of adaptation of the geometry of the cross-section and uptake parameters to the incoming~MBF}.
Large values of~$\gamma$~($\gamma\gg1$) mean that oxygen is quickly transferred to the fetal circulation at the beginning of the stream-tube and is rapidly depleted, so that poor in oxygen maternal blood flows through the remaining part.
Thus, this remaining part does not function efficiently.
Small values of~$\gamma$~($\gamma\ll1$) mean that maternal blood passes too quickly through the placenta as compared to the oxygen extraction time, so that a considerable part of the incoming oxygen flow may not be transferred. One can then speculate that transport of oxygen is the most efficient in the placentas, for which~$\gamma$ is of the order of~1. $\gamma\approx1.4$ calculated from the model parameters suggests that a healthy placenta may indeed function optimally.

\begin{figure*}[tb]
\newcommand{\picHeight}{0.25}  
\vspace{-4ex} \centering 
\subfloat[]{
\includegraphics[height=\picHeight\paperwidth, natheight=477, natwidth=912]
{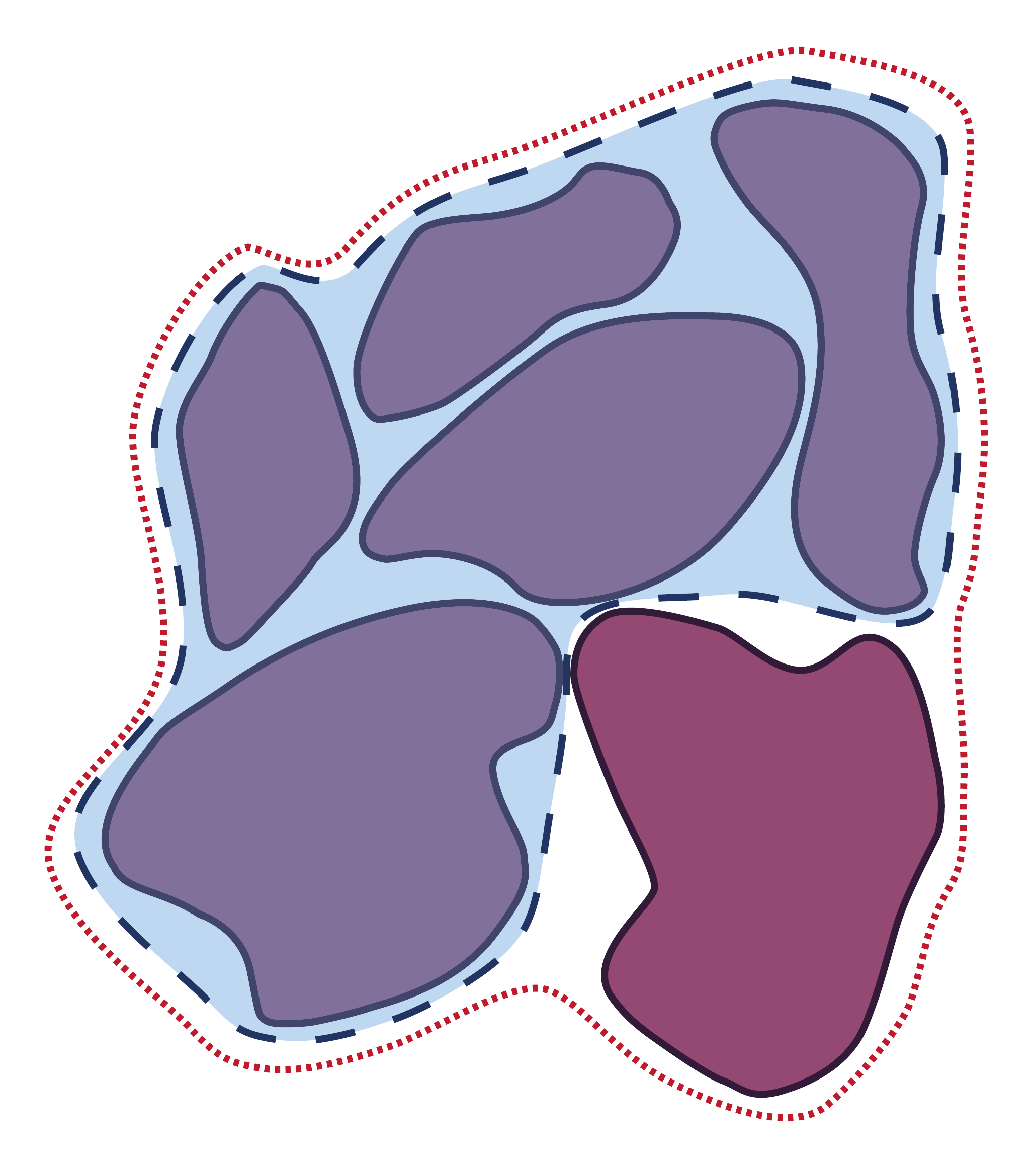}
\label{fig:DualityScheme} }
\hspace{1cm}
\subfloat[]{
\includegraphics[height=\picHeight\paperwidth, natheight=477, natwidth=912]
{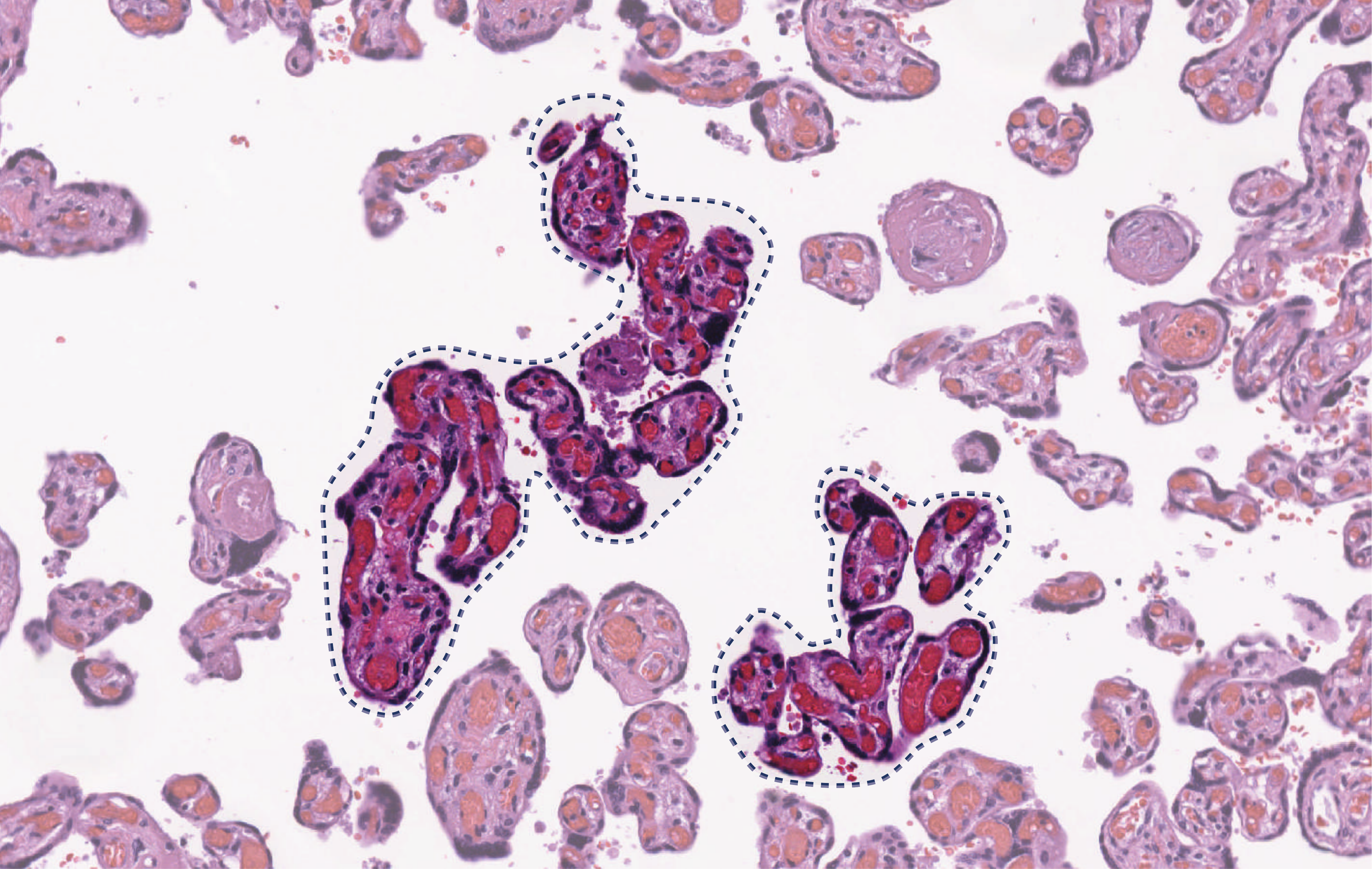}
\label{fig:DualityExample} }

\caption{
Illustration of the difference between the total villous perimeter and the effective absorbing villous perimeter.
\protectedSubref{fig:DualityScheme}
Example of an isolated group of villi. In the shaded villi group, some parts of absorbing villi boundaries are inefficient as they are screened by other villi from the outside of the isolated group, where the main reservoir of oxygen is supposed to be. The effective absorbing perimeter of the group is close to the perimeter outlined by the dashed line. This perimeter of the~IVS surrounding the villi can be several times smaller than the total perimeter of the villi in the shaded group. Note also that adding a new villus into such group~(the villus outside the shaded group) does not increase the effective absorbing perimeter proportionally to the increase of the total villous perimeter. In the example in the figure, the new perimeter~(the dotted contour) has approximately the same length as the old one.
\protectedSubref{fig:DualityExample}
Illustration of the same concept in a histological slide of the human placenta. Two dashed contours show the effective absorbing perimeters of two groups of villi, which are considerably smaller than the total perimeters of villi in the groups.
Discussion of similar screening concepts can be found in~\citet{Sapoval2002,Felici2005,Gill2011}
}
\label{fig:DualityIllustration}
\end{figure*}

\subsection{The analytical theory}

The advantages of the analytical solution over the numerical one are numerous:
\begin{enumerate}

\item
Oxygen uptake can be estimated for a histological cross-section of arbitrary geometry.

\item 
The villi density $\phi$ and the effective villi radius $r_\mathrm{e}$ are shown to be the only geometrical parameters necessary to predict oxygen uptake of a rather uniform villi distribution in a placental cross-section~(see Figs~\ref{fig:UptakeResults}, \ref{fig:ResultsDependenceOnVelocity}). These two parameters allow for a simple application of the theory to distributions of villi of arbitrary shapes. The validity of the theory in the case of strongly non-uniform villi distributions remains to be investigated.

Finer details of villi distributions which produce differences between numerical and analytical results in Figs~\ref{fig:UptakeResults} and~\ref{fig:ResultsDependenceOnVelocity}, are \enquote{stored} in the coefficient~$\kappa$. This coefficient encompasses not only the details of villi distributions, but also determines the strength of their influence on oxygen uptake at a given length~$L$. In other words, it quantitatively describes the fact that in each geometry, different regions of the~IVS are not equivalent due to random distribution of villi, and that with length~$L$, oxygen in some regions is exhausted faster than in other regions. 
However, the dependence of~$\kappa$ and~$\alpha$ on~$L$ is rather weak as can be seen in Fig.~\ref{fig:UptakeLVilliDensity}, in which~$\FApprox(L)$ is plotted for all lengths with the same~$\kappa=\kappa(L_0)\approx0.35$. In the first approximation,~$\kappa$ can hence be considered independent of~$L$.

\item
The efficiency of oxygen transport in a given placental cross-section can be estimated by means of \emph{oxygen extraction efficiency}~$\zeta$ and \emph{villi density efficiency}~$\etaE$ plotted in Fig.~\ref{fig:efficiency_diagrams}. %
For these two quantities, simple analytical formulas and diagrams are provided, which allow for comparison of different placentas or placental regions once the parameters~$\phi$ and~$\gamma$ are calculated for them.
To have predictive power, the efficiency estimations provided by the model need to be studied for correlations with independent indicators of placental exchange efficiency, such as placenta shape, placenta weight, placenta-fetus birth-weight ratio~\citep{Misra2010, Hutcheon2012} or pulsatility indices determined by Doppler velocimetry in the umbilical cord or maternal vessels~\citep{Todros1999, Madazli2003}, which were demonstrated to vary between normal and pre-eclamptic pregnancies or pregnancies complicated by fetal growth restriction.

\item
The analytical theory suggests that oxygen uptake in the human placenta is rather robust to changes of villi density. Indeed, the diagram in Fig.~\ref{fig:efficiency_abs} shows that placental villi density can vary by about~\unit[10]{\%} around the optimal value with the villi density efficiency~$\etaE$ staying in the~\unit[90\text{--}100]{\%} interval. Far from the optimal villi density,~$\etaE$ tends to decrease faster.


\item 
One can analyze the consequences of neglecting oxygen-hemoglobin reaction on the predictions of oxygen uptake and the optimal villi density of a placental region. Moreover, the theory gives a method of recalculation of the results obtained for no-Hb blood in artificial placenta perfusion experiments into those for normal blood. Imagine that at the end of an artificial perfusion experiment with no-Hb blood, one obtains the total oxygen inflow~$\FInPerf$ into the placenta, fetal oxygen uptake~$\FPerf$ and the average villi density~$\phi$ from histomorphometry of the same placenta~(note that~$\FInPerf$ and~$\FPerf$ differ from~$\FIn$ and~$F$ which would have been obtained for normal blood). From these data one can calculate~$\FZeroPerf=\FInPerf/(1-\phi)$~(see the discussion of Eq.~\fm{fm:GammaAndF0Definitions}) and then~$\gammaPerf$ as a root of Eq.~\fm{fm:FinalTheoreticalApproximationEpsilon}~(with~$\FApprox$ replaced by~$\FPerf$).
These values can be recalculated for blood containing~Hb:~$\gamma=\gammaPerf/B$ and~$F_0=\FZeroPerf B$, where~$B\approx94$~(Table~\ref{tblCalculationsParameters}), and can be substituted into Eq.~\fm{fm:FinalTheoreticalApproximationEpsilon} to give oxygen uptake~$F$ in the same placenta for blood containing~Hb.
One can see that oxygen uptake in a no-Hb perfusion experiment gives on the average a hundred-times underestimation of the real uptake.
Finally, the values of~$\gamma$ and~$\phi$ for the given placenta can be compared with the diagram in Fig.~\ref{fig:villi_density_optimality_diagram} to determine how far the geometry of the region is from the optimal one.
Note that this recalculation introduces a small error as in no-Hb case the diffusive part of the total flow, which is omitted in Eq.~\fm{fmFInftyTwoComponents}, becomes important;

\item 
The computation time is reduced since calculations of eigenfunctions and eigenvalues of the diffusion equation are not required. Note that due to long computation time, numerical simulations of~\citet{Serov2015Optimality} had to be performed on a smaller placentone radius~$\RNum$ and then rescaled to the radius~$R$ by multiplying oxygen uptake by~$R^2/\RNumSquared$. This constraint does not apply to the analytical theory. In particular, good agreement between both approaches justifies the rescaling of results performed in the numerical calculations.

\end{enumerate}

Note finally that the derivation of Eq.~\fm{fm:GammaAndF0Definitions} implies that, strictly speaking,~$P$ is  not the total perimeter of the villi, but the effective absorbing perimeter of the villi~(i.e. only its part that is directly in contact with the~IVS). 
In the case of well-separated villi, there is no difference between the two definitions. 
However, it is not always the case in placental cross-sections. 
For instance, in Fig.~\ref{fig:DualityExample} one can see several isolated groups of villi, inside which villi lie so close to each other, that there is virtually no~IVS left between them. Parts of the villous boundary which are not in contact with large parts of the~IVS are then \emph{screened} from participating in oxygen uptake, and hence should not be accounted for in the effective absorbing perimeter of the villi. This remark can be understood by considering the fact that oxygen diffuses in the~IVS, and only parts of the villous boundary that are in contact with the~IVS will participate in the uptake. %
In the case of well separated singular villi, the entire perimeter is absorbing. 
A schematic description of this situation is shown in Fig.~\ref{fig:DualityScheme}.
Note finally that in our model the screening effect is implicitly taken into account by the diffusion equation~\fm{fmEigenValueEquation01}.

\section{Conclusions}

In the present work, an analytical solution to the diffusion-convection equation governing oxygen transport in the human placenta has been developed.
Oxygen uptake was calculated for an arbitrary cross-sectional geometry of the stream tubes of maternal blood. %
It was shown that for a rather uniform spatial distribution of villi in a placental cross-section, only two geometrical characteristics, villi density~$\phi$ and the effective villi radius~$\rEff$, are needed to predict fetal oxygen uptake.

It was also demonstrated that all the parameters of the model do not influence oxygen uptake independently, but instead form two combinations:~(i) the maximal oxygen inflow of one placentone~$F_0$, and~(ii) the ratio~$\gamma$ of the transit time of maternal blood through the~IVS and the oxygen extraction time. These two parameters together with villi density determine oxygen uptake. 
Analytical formulas and diagrams were obtained which allow for oxygen uptake calculation and quantitative estimation of the efficiency of oxygen transport of a given placental region based on measurements of~$\phi$ and~$\rEff$.

Finally, a fictitious case of blood containing no hemoglobin was analyzed to study oxygen transport in artificial placenta perfusion experiments. %
It was demonstrated that artificial perfusion experiments with no hemoglobin tend to give a two-orders-of-magnitude underestimation of the \emph{in vivo} oxygen uptake. %
A method of recalculation of the results of artificial perfusion experiments to account for oxygen-hemoglobin dissociation was proposed.

Once combined with image analysis techniques, the proposed analytical theory can be the mathematical ground for a future tool of fast diagnostics of placenta efficiency based on histological placental slides.

\section{Acknowledgements}

The authors thank Dr~Paul Brownbill for fruitful discussions.

This study was funded by the International Relations Department of Ecole Polytechnique as a part of the Ph.D. project of A.S.~Serov, by Placental Analytics~LLC,~NY, by SAMOVAR project of the Agence Nationale de la Recherche n$^\circ$ 2010-BLAN-1119-05 and by Agence Nationale de la Recherche project~ANR-13-JSV5-0006-01.

\bibliographystyle{my_plainnat}
\footnotesize
\setlength{\bibsep}{0pt}
\bibliography{library}

\end{document}